\documentclass[a4paper,fleqn]{cas-sc}

\usepackage[authoryear]{natbib}
\usepackage{multirow}
\usepackage{graphicx}
\usepackage{subfigure}

\begin{document}
\let\WriteBookmarks\relax
\def\floatpagepagefraction{1}
\def\textpagefraction{.001}
\shorttitle{}
\shortauthors{Yong Li et~al.}
\let\printorcid\relax

\title [mode = title]{Monitoring urban ecosystem service value using dynamic multi-level grids}                      

\author[1]{Zhenfeng Shao}

\author[1]{Yong Li}
\author[3]{Xiao Huang}
\author[2]{Bowen Cai}
\author[2]{Lin Ding}
\author[1]{WenKang Pan}
\author[1]{Ya Zhang}
\address[1]{State Key Laboratory for Information Engineering in Surveying, Mapping and Remote Sensing, Wuhan University, Wuhan 430079, China}
\address[2]{School of Remote Sensing and Information Engineering, Wuhan University, Wuhan 430079, China}
\address[3]{Department of Geosciences, University of Arkansas, Fayetteville, AR 72701, USA}

\begin{abstract}
Ecosystem services are the direct and indirect contributions of an ecosystem to human well-being and survival. Ecosystem valuation is a method of assigning a monetary value to an ecosystem with its goods and services, often referred to as ecosystem service value (ESV). With the rapid expansion of cities, a mismatch occurs between urban development and ecological development, and it is increasingly urgent to establish a valid ecological assessment method. In this study, we propose an ecological evaluation standard framework by designing an ESV monitoring workflow based on the establishment of multi-level grids. The proposed method is able to capture multi-scale features, facilitates multi-level spatial expression, and can effectively reveal the spatial heterogeneity of ESV. Taking Haian city in the Jiangsu province as the study case, we implemented the proposed dynamic multi-level grids-based (DMLG) to calculate its urban ESV in 2016 and 2019. We found that the Haian city’s ESV showed considerable growth (increased by 24.54 million RMB). Negative ESVs are concentrated in the central city, which presented a rapid trend of outward expansion. The results illustrated that the ongoing urban expanse does not reduce the ecological value in the study area. The proposed unified grid framework can be applied to other geographical regions and is expected to benefit future studies in ecosystem service evaluation in terms of capture multi-level spatial heterogeneity. 

\end{abstract}

\begin{graphicalabstract}
	\includegraphics[scale=0.28]{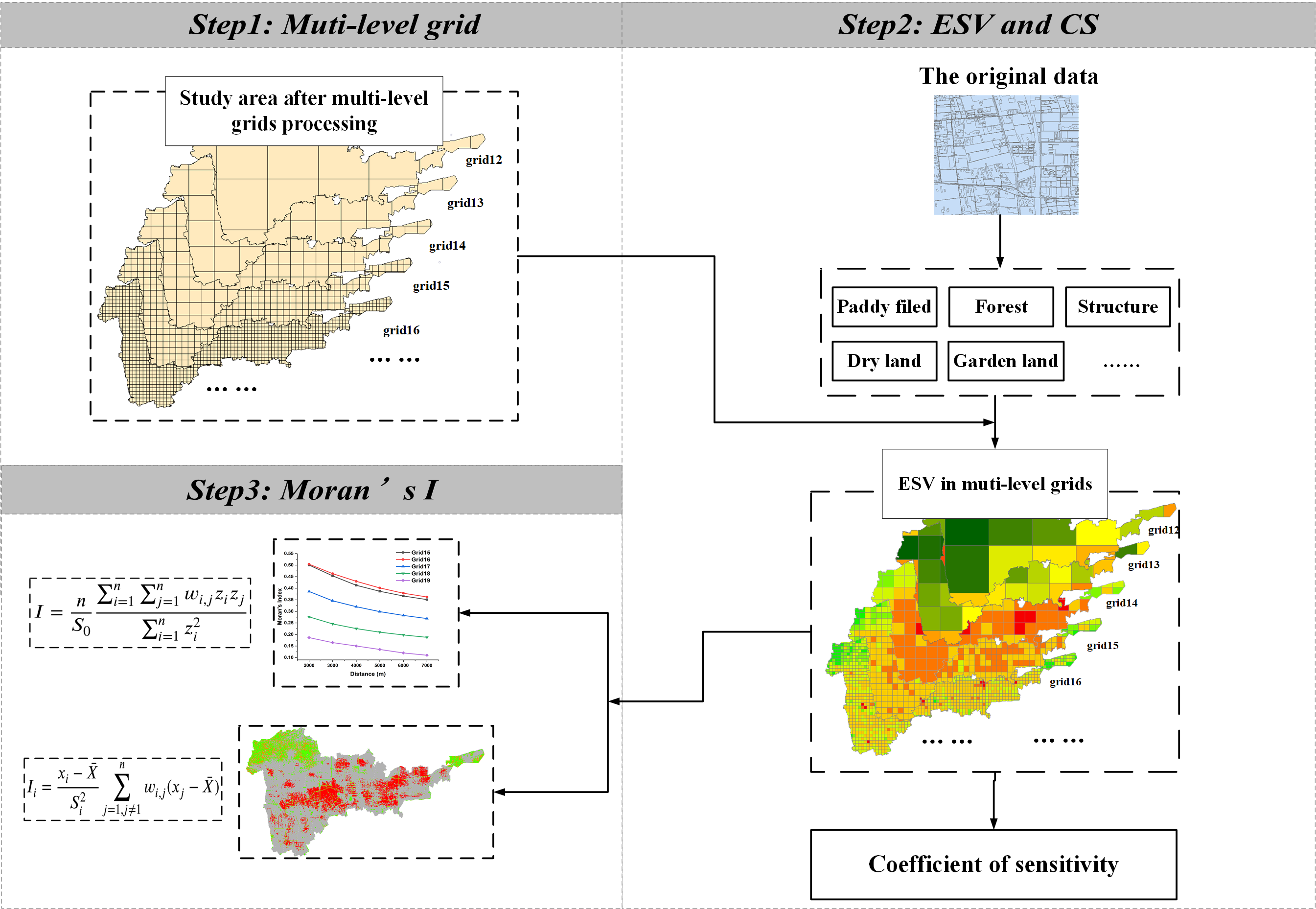}
\end{graphicalabstract}

\begin{highlights}
\item We propose a dynamic multi-level grid-based ESV model.
\item The proposed model with multi-level functions mitigates the limitations of the existing macro-scale models.
\item The model with dynamic multi-level grids is expected to benefit better monitoring of fine-grained urban ecology.  
\end{highlights}

\begin{keywords}
Ecosystem services  \sep  Dynamic multi-level  grids  \sep ESV \sep Land use/cover change \sep
\end{keywords}

\maketitle

\section{Introduction}

Ecosystem services are defined as the benefits or contributions provided by natural ecosystems 
that can be used by humans \citep{costanzaValuingEcologicalSystems2011}.
Ecosystems are natural resources and assets, serving as the basis of human survival and development \citep{BRAAT20124,BORGES2021107458,HUANCAPAZAHILASACA2021107316}. 
Natural ecosystems provide the material basis and ecological services for the survival and development of human society, and a healthy ecosystem can ensure the sustainability of human development \citep{pengResearchProgressEvaluation2011}.
However, there exists a paradox at our time: while we are obtaining better knowledge that facilitates the assessment of ecosystems' states, the health of ecosystems continues to plummet \citep{RAPPORT201310,LIN2021107461}.
Sustained population growth, industrialization, and urbanization are attaching a considerable strain on ecosystems around the world, increasing the likelihood that the available ecosystem services are not able to withstand such strain \citep{RASOOL2021107447, WANG2021107449}.
Ecological-environmental effects, caused by a variety of issues such as over-exploitation of resources, climate change, land degradation, biodiversity decline, and environmental pollution, lead to loss and degradation of ecosystem service functions, threatening human safety, public health, as well as regional and global ecological security \citep{kumar2010economics,TAPOLCZAI2021107322,CHEN2021107448}.
Thus, protecting and enhancing ecosystem services has become an urgent challenge, and monitoring the health of ecosystems is one of the essential steps to achieving sustainability. 

After the theoretical and methodological evaluation of ecosystem services proposed and developed by \citet{costanzaValueWorldEcosystem1997}, it has attracted wide attention. The Millennium Ecosystem Assessment (MA) project has further promoted the progress of ecosystem services and gradually become one of the research hotspots in the fields of ecology, ecological economics, and geography.

Land cover changes, containing huge ecological and economic benefits and resulting from social and economic development, are responsible for the dynamics in many natural phenomena and ecological processes on the surface of the Earth. Studies that target land cover patterns and their impact on the ecological environment are of great significance to promote the coordinated development of the regional economy and environment. Strengthening the quantitative analysis and evaluation of regional Ecosystem Service Value (ESV) facilitates the understanding of the relationship between ecosystem service value and human well-being.

Recent years have seen a growing literature on ecosystem service valuation (ESV). Generally, two types of monetary valuation approaches are widely applied for ESV \citep{xieDynamicChangesValue2017};
one is referred to as primary data based approach \citep{CBA:534824},
and the other is referred to as unit value-based approach \citep{costanzaChangesGlobalValue2014,xieDynamicChangesValue2017}.
\citet{xieDynamicChangesValue2017} modified the value coefficients of Chinese ecosystems based on Costanza's parameters \citep{costanzaValueWorldEcosystem1997}.
\citet{TIANHONG20101427} used Shenzhen as an example to analyze the economic value of urban ecosystem services and estimate the changes in ecosystem services value caused by changes in land use during the urbanization process.
\citet{nunezForestsWaterValue2006a} estimated the economic value of Chile’s temperate forests, whose main function is to maintain freshwater supplies.
\citet{heinSpatialScalesStakeholders2006} analyzed the spatial scale of ecosystem services and studied how stakeholders on different spatial scales assign different values to ecosystem services.
\citet{sannigrahiEcosystemServiceValue2019} from a spatiotemporal perspective, measured ESV of 17 key Ecosystem services of Sundarbans Biosphere Reserve in India using remote sensing data.
\citet{yongxiuSpatiotemporalVariationsCoupling2020}comprehensively evaluate the Qinghai-Tibet Plateau (QTP) county-level ecology based on the four-quadrant model on the Qinghai-Tibet Plateau using data from a variety of sources. 
Due to the large spatial heterogeneity of QTP, however, other human factors may also affect ecological quality, e.g., the development of mineral resources, traffic volume, tourism, pollution, dams, ecological restoration projects, to list a few.

Many scholars have conducted theoretical and empirical studies on the impact of land use on ecosystem service value. Despite these efforts, they are limited to the analysis of Ecosystem Service Value (ESV) at a single scale in a city \citep{TIAN2020106543,zhangExploringOptimalIntegration2018}, lacking the evolution analysis of multiple scales. In addition, given the fact that buildings play an essential role in urban fabrics, more attention in the role of buildings should be paid to the study of urban ESV. In this study, we aim to provide a new perspective for the method of evaluating ESV by proposing a multi-scale assessment method of urban ecosystem health based on the framework of ecosystem vitality, organization, resilience, and service, which can be applied to assess the ecological service in other cities beyond the study area. Taking Haian City, Jiangsu Province, China, as a study case, we build a comprehensive index system using dynamic multi-level grids for evaluating the urban ecosystem.

\section{Materials and Methods}
\subsection{Study Area}
\begin{figure}[ht]
	\centering
	\includegraphics[scale=0.6]{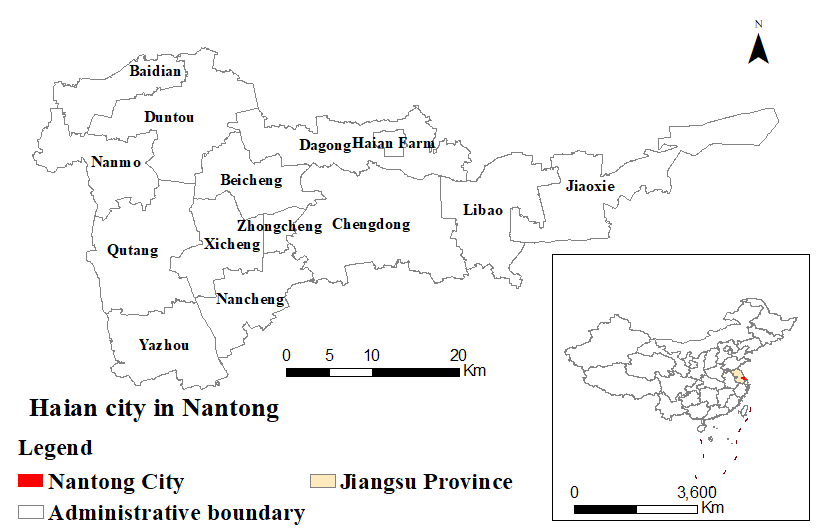}
	\caption{ Location of Haian City.}
	\label{fig:studyarea}
\end{figure}

Haian city is located in the southeast of Jiangsu Province, China. It is on the north bank of the lower Reaches of the Yangtze River, adjacent to the Yellow Sea in the east and the Yangtze River in the south. Haian city enjoys a superior geographical position and is one of the famous port cities in China, As shown in the figure \ref{fig:studyarea}.
Since the reform and opening up, rapid economic development and continuous improvement of urbanization have led to a large number of cultivated land being occupied by construction land, land cover and urban landscape have changed greatly, and a series of ecological environmental problems have emerged.
By studying the influence of urban land use change on ecosystem service function, this paper provides decision support for sustainable land resource utilization and ecological environment protection in Haian city.

Haian city is located in the southeast of Jiangsu Province, China. It is on the north bank of the lower Reaches of the Yangtze River, adjacent to the Yellow Sea in the east and the Yangtze River in the south (Figure \ref{fig:studyarea}). Haian city enjoys a superior geographical position and is one of the famous port cities in China. Since the Economic Reform and open up policy, rapid economic development and continuous improvement of urbanization have led to a large amount of cultivated land being occupied by construction land. The land cover and urban landscape in Haian city have changed dramatically with the emergence of a series of ecological, environmental problems. By investigating the influence of urban land use changes on ecosystem service function, we aim to provide decision support for sustainable land resource utilization and ecological environment protection in Haian city.

\subsection{Data}
The 2016 and 2019 datasets of geographical conditions of Haian city used in this paper are from the Nantong Institute of Surveying and Mapping, a research institution in Nantong city. These data sets are mainly derived through the interpretation of remote sensing images, assisted with ground surveys and existing land use maps. The classified land covers in the datasets well reflect the natural properties or conditions of the land surface natural creations and artificial constructions.
The classified land covers in the datasets well reflect the natural properties or conditions of the land surface natural creations and artificial constructions.

The data collection was completed independently by a professional team, adopting the mode of unified standard, direct data reporting, centralized database building, strict production organization, and strict quality control measures to ensure that the spatial location and attribute information of the census elements were consistent. The comprehensive data system contains the classification, area, and spatial distribution of 146 categories of surface elements in the form of points, lines, planes, and geographical entities from the aspects of topography, landform, land cover, important geographical conditions, etc. 
According to administrative divisions, elevations, slopes, and other units, diversified statistical analysis can be carried out in three-dimensional space to reflect the distributive characteristics of geographical conditions in detail. In the process of data formation, the same data interpretation method and the same classification categories are adopted. 
The land use categories of the original data include a total of 10 first-grade categories (e.g., cultivated land, woodland, grassland, and etc.), 45 second-grade categories, and 87 third-grade categories. Among them, the proportion of dry land and paddy field are dominant categories. To highlight the wide distribution of paddy fields in Haian and facilitate the identification of the vice ecological value of construction land, the dataset was reclassified into 9 categories in this study (Table \ref{landClass}), including paddy field, dry land, garden land, woodland, grassland, house building, structure, artificial pile digging land and water bodies. The land use map is then edited, calibrated, and coded in the ArcMap10.2 software for the following ecosystem service value calculation and spatial analysis.

\begin{table}[]
	\centering
	\caption{Land over classification table.}
	\label{landClass}
	\resizebox{\textwidth}{!}{%
		\begin{tabular}{ll}
		\hline
			Land over types &
			Description \\ \hline
			Paddy field &
			\begin{tabular}[c]{@{}l@{}}
				 \quad Refers to cultivated land that has water source guarantees and irrigation\\ 
				facilities, and can be irrigated normally in normal years, for the cultivation\\ 
				of aquatic crops such as rice and lotus roots, including cultivated land where\\ 
				rice and dry land crops are rotated.
			\end{tabular} \\ 
			Dry land &
			\begin{tabular}[c]{@{}l@{}}
				\quad Refers to cultivated land without irrigation water sources and facilities that\\ 
				rely on natural precipitation to grow crops; dry crop cultivated land with water\\
				sources and irrigation facilities that can be irrigated normally in normal years;\\ 
				cultivated land mainly for vegetable cultivation; fallow land and rotation of\\
				normal cropping Rest.
			\end{tabular} \\ 
			Garden land &
			\begin{tabular}[c]{@{}l@{}}
				\quad Refers to continuous artificial planting, perennial woody, herb crops, intensive\\ 
				management, including Joe irrigation orchard, rattan orchard, tea garden, mulberry\\
				garden, nursery, and flower garden.
			\end{tabular} \\ 
			Forest &
			\begin{tabular}[c]{@{}l@{}}
				\quad Refers to the land surface covered by patches of natural forests, secondary forests\\
				 and artificial forests, including trees, shrubs, bamboo and other types.
			 \end{tabular} \\ 
			Grassland &
			\begin{tabular}[c]{@{}l@{}}
				\quad Refers to the land surface covered mainly by herbaceous plants, including natural\\
				grassland and artificial grassland.
			\end{tabular} \\ 
			Settlement &
			    \quad Includes housing construction area and independent housing construction. \\ 
			Structure &
			\begin{tabular}[c]{@{}l@{}}
				\quad Refers to the engineering entity or ancillary building facilities, including\\
				hardened surface, hydraulic facilities, greenhouses, curing ponds, industrial\\
				facilities, etc.\end{tabular} \\ 
			\begin{tabular}[c]{@{}l@{}}Artificial pile \\ digging land\end{tabular} &
			\begin{tabular}[c]{@{}l@{}}
				\quad Refers to a surface that is permanently covered by abandoned objects formed \\by
				human activities or that has been artificially excavated and exposed as a result of\\ 
				large-scale civil works, including open excavation sites, piles, construction sites,\\
				 and other artificial excavation sites.
			\end{tabular} \\ 
			Water area &
			\begin{tabular}[c]{@{}l@{}}
				\quad Refers to natural land waters and land used for water conservancy facilities,\\  mainly lakes and rivers.\end{tabular} \\ \hline
		\end{tabular}%
	}
\end{table}

\subsection{Methods}

\begin{figure}[htbp]
	\centering
	\includegraphics[scale=0.28]{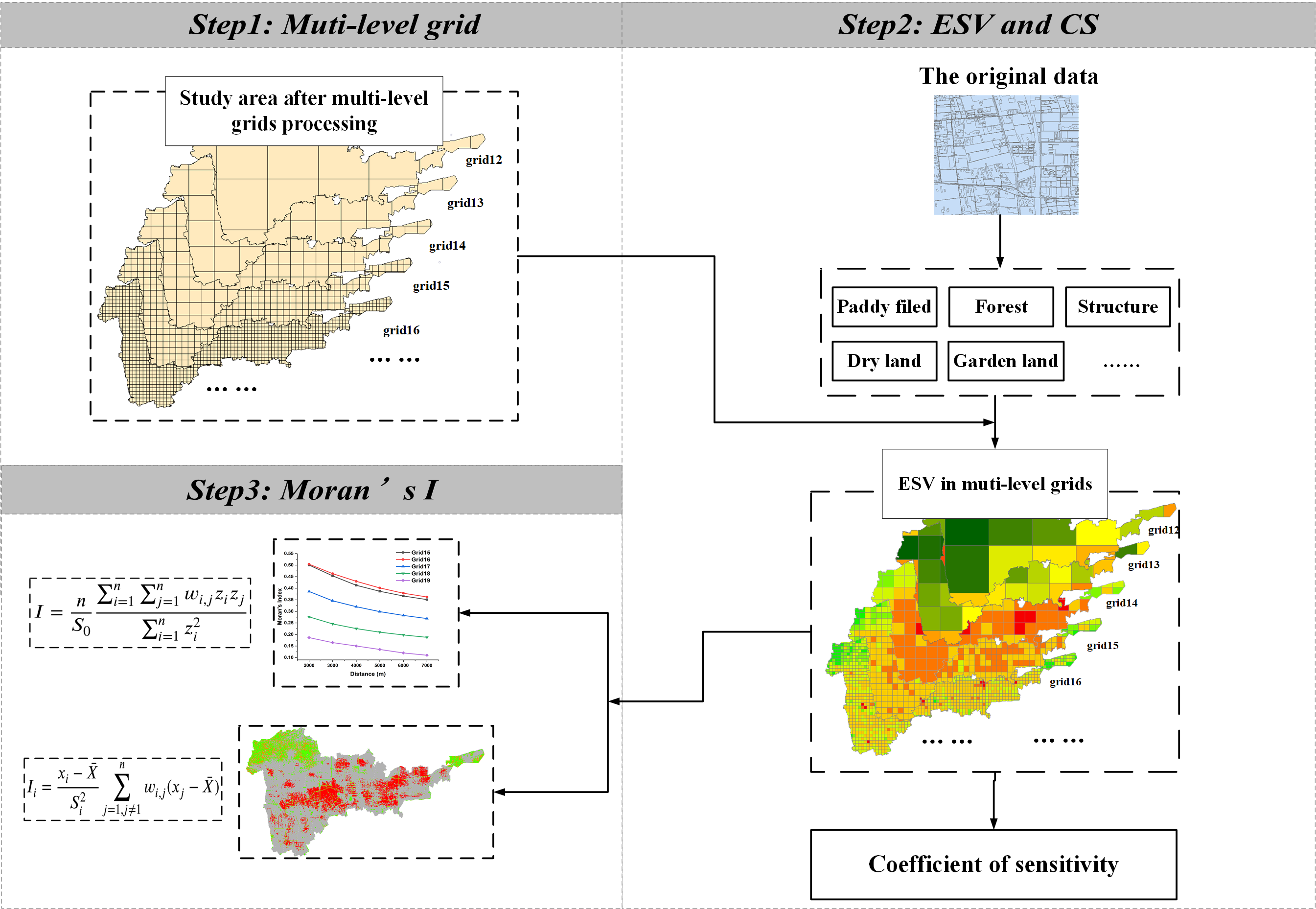}
	\caption{Technical processes for evaluating dynamic multi-level grids-based model}
	\label{fig:method }
\end{figure}

The multi-level model we proposed contains three steps (\ref{fig:method }): 1) generating multi-level grids data; 2) processing raw data and calculating ESV and CS; 3) calculating The Global Moran's Index and The Local Moran's Index.

\subsubsection{Ecosystem service value assessment}

In this paper, non-construction land includes paddy field, dry land, garden land, woodland, grassland, and water bodies. We adopted the ESV evaluation model from \citet{costanzaValueWorldEcosystem1997} for evaluation:
\begin{equation}\label{ESV}
	ESV_{n}=\sum_{i}\sum_{j}A_{j} \times VC_{ij}
\end{equation}
where $ESV_{n}$ is the ecosystem service value of non-construction land;
$A_{j} $ is the area of the j class land;
$VC_{ij}$ is the unit area value of the ecological service in item $i$ of type $j$ class land.

Ecosystem services are derived from the flows of materials, energy, and information through a certain ecosystem determined by its structure and processes. Given different ecological functions, ecosystem services were divided into four primary categories, i.e., provisioning services, regulating services, habitat services, and cultural services \citep{kumar2010economics}. 
After that, these services were further divided into 11 secondary services, ie, food, materials, water, air quality regulation, climate regulation, waste treatment, regulation of water flows, erosion prevention, maintenance of soil fertility, habitat services, cultural and amenity services.
Given the spatial heterogeneity of ecosystem services, we follow the approach from \citet{xieDynamicChangesValue2017} to modify the value coefficient of global terrestrial ecosystem services calculated by \citet{costanzaValueWorldEcosystem1997}, obtained through the investigation of 200 ecologists in China.
The equivalent cofficients table for ecosystem service value combines the vegetation and land type characteristics of Haian based on the revised results of \citep{xieDynamicChangesValue2017}.

We follow the methods from \citet{wanEffectsUrbanizationEcosystem2015} and \citet{xueDynamicVariationsEcosystem2015} to calculate the value of gas regulation, water supply, and environmental purification functions, and use the market value method to calculate the value of cultural and amenity services. 

(1)The ecosystem service value of gas adjustment for construction
land was calculated as:
\begin{equation}\label{Eq}
	E_{q}=-\frac{C_{q}}{S}
\end{equation}
where $E_{q}$ is ecosystem service value of gas adjustment unit area for
construction land; $C_{q}$ is the cost for dealing with waste gas; $S$ is the total area of construction land.

(2)The Ecosystem service value of hydrological adjustment for construction land was calculated as:
\begin{equation}\label{CVw}
	CV_{w}=-\frac{C_{w}+WP_{w}}{A_{c}}
\end{equation}
where $CV_{w}$ is ecosystem service value of hydrological adjustment unit area for construction land; 
$C_{w}$ is the cost for dealing with waste water; 
$W$ is the total quantity of water consumption, 
$P_{w}$ is the unit price of water consumption, and 
$A_{c}$ is the total area of construction land.

(3)The ecosystem service value of waste treatment for construction land was calculated as:
\begin{equation}\label{CVc}
	CV_{c}=-\frac{C_{r}+C_{c}}{A_{c}}
\end{equation}
where $CV_{c}$ is the ecological service value of environmental purification per unit area of construction land;
$C_{r}$ is the cost of household waste treatment;
$C_{c}$ is the cost of industrial waste treatment;
$A_{c}$ is the total area of the construction land.

(4)The ecosystem service value of cultural and amenity services for construction land was calculated as:
\begin{equation}\label{CVa}
	CV_{a}=\frac{C_{t}}{A_{c}}
\end{equation}
where $CV_{a}$ is the ecological service value of the aesthetic landscape per unit area of the construction land;
$C_{t}$ is the revenue of tourist attractions;
$A_{c}$ is the total area of the construction land.

\begin{table}[htbp]
	\centering
	\caption{Annual average ecosystem service value of the unit area of different land cover type in Haian (RMB/ha).}
	\resizebox{\textwidth}{!}{
	\begin{tabular}{llllllllllll}
	\hline
	\multirow{2}{*}{\begin{tabular}[c]{@{}l@{}}Ecosystem\\ classification\end{tabular}} &
	\multicolumn{3}{l}{Provisioning services} &
	\multicolumn{4}{l}{Regulating services} &
	\multicolumn{3}{l}{Habitat Services} &
	Cultural \\ \cline{2-12} 
	&	Food &	Materials & 	Water &
	\begin{tabular}[c]{@{}l@{}}Air quality\\ regulation\end{tabular} &
	\begin{tabular}[c]{@{}l@{}}Climate   \\ regulation\end{tabular} &
	\begin{tabular}[c]{@{}l@{}}Waste   \\ treatment\end{tabular} &
	\begin{tabular}[c]{@{}l@{}}Regulation   \\ of water \\ flows\end{tabular} &
	\begin{tabular}[c]{@{}l@{}}Erosion   \\ prevention\end{tabular} &
	\begin{tabular}[c]{@{}l@{}}Maintenance   \\ of soil \\ fertility\end{tabular} &
	\begin{tabular}[c]{@{}l@{}}Habitat\\ Services\end{tabular} &
	\begin{tabular}[c]{@{}l@{}}Cultural\& \\ amenity \\ services\end{tabular} \\ \hline
	Paddy field &  2029.12 &	134.28 &	-3923.96 &	1656.12 &	850.44  &	253.64  &	4058.24 &	1536.76 &	179.04 &	193.96  &	89.52 \\ 
	Dry land    &	1268.2 &	596.8  &	29.84    &	999.64  &	537.12  &	149.2   &	402.84  &	1536.76 &	179.04 &	193.96  &	89.52 \\ 
	Garden land &	283.48 &	641.56 &	328.24   &	1700.88 &	6311.16 &	1909.76 &	4998.2  &	2566.24 &	193.96 &	2342.44 &	1029.48 \\ 
	Forest      &	432.68 &	984.72 &	507.28   &	3237.64 &	9698    &	2879.56 &	7072.08 &	3953.8  &	298.4  &	3580.8  &	1581.52 \\ 
	Grassland   &	566.96 &	835.52 &	462.52   &	2939.24 &	7773.32 &	2566.24 &	5699.44 &	3580.8  &	268.56 &	3252.56 &	1432.32 \\ 
	Settlement  &	   0   &	0      &	-27552   &	0       &	-6299   &-6495.72   &	0       &	0       &	0      &	0       &	4323.08 \\ 
	Structure   &	0      &	0      &	0        &	0       &	0       &	5000    &	0       &	0       &	0      &	0       &	1000 \\ 
	\begin{tabular}[c]{@{}l@{}}Artificial pile \\ digging land\end{tabular}
	           &	0      &	0       &	0        &	-50     &	0       &	-100    &	0       &	0        &	0      &	0       &	-500 \\ 
	Water area &	1193.6 &	343.16 &	12368.68 &	1148.84 &	3416.68 &	8280.6 &	152542.1 &	1387.56 &	104.44 &	3804.6 &	2819.88 \\ \hline
    \end{tabular}
    }

\end{table}

\subsubsection{Coefficient of sensitivity for ecosystem service value (CS)}
The change of ecosystem service value caused by the change of equivalent factor (namely the sensitivity of ecosystem service value to equivalent factor) is called the coefficient of sensitivity (CS).
The purpose of calculating CS is to verify the representation of ecosystem types and the accuracy of the ecological value coefficient in various land cover classifications.
CS was calculated using the standard economic concept of elasticity as follows \citep{kreuterChangeEcosystemService2001,sunSpatiotemporalEvolutionScenarios2019}:
\begin{equation}\label{CS}
	CS=\frac{(ESV_{n}-ESV_{m})/ESV_{m}}{(VC_{ni}-VC_{mi})/VC_{mi}}
\end{equation}
Where $CS$ represents the coefficient of sensitivity for ecosystem service value;
$ESV$ is the total ecological service value, $VC$ is the value coefficient, $m$ and $n$ are the initial and adjusted states, respectively;
$i$ is the land use type.

During the calculation, the percentage changes of total ESV for a given percentage change in a value coefficient and CS from a 50\% adjustment were calculated.
When CS > 1, the estimated ESV is considered to be elastic with respect to the value coefficient; 
while when CS < 1, otherwise.

\subsubsection{Ecosystem service value assessment based on multi-level grids}

Geographic multi-level grids are generally adopted to assist the organization and storage of geospatial data. We use the geographic grids from GeoSot \citep{ijgi5090161,shaoSpatialInformationMultigrid2005}, with each grid having a unique identity. We take advantage of the multi-level features of the GeoSot grids to establish an ecological assessment framework. Under this unified grid framework, we hope to investigate the changes of ecological indicators in different cities and regions around the world under different grid schemes. 

The grid generation approach adopted in this study can be divided into the following two major steps. First, the quadtree recursion is used to divide the earth’s surface with a total of 32 levels (the scaling ratio between the upper and lower levels is $\frac{1}{4}$). Second, the grid coding scheme maintains the structure of "degree, minute, and second". To facilitate the efficient operation among grids, coordinates (latitude and longitude) were expanded. Such a coding method can be compatible with traditional map sheet of surveying and mapping standard, achieving an efficient multi-level analysis and laying a foundation for the effective application of global ecological monitoring and evaluation.
Existing efforts in calculating  ESV usually consider only a fixed spatial unit, falling short to comprehensively analyze the changes of urban ESV from multiple spatial perspectives. In this study, we established grids of 14-19 levels, with each grid having a unique identity. Table \ref{gridscale} shows the side length and number of grids at the level of 14-19 for Haian City in our study area.
We use multi-level grids to calculate the ESV of each grid cell at different grid levels:

\begin{equation}
	\label{ESVg}
	ESV_{g}=\sum_{i}\sum_{j}A_{jg} \times VC_{ij}
\end{equation}
where $ESV_{g}$ is the ecosystem service value of land over types in a grid cell;
$A_{jg} $ is the area of the $j$ class land in a cell grid;
$VC_{ij}$ is the unit area value of the ecological service in item $i$ of type $j$ class land.

\begin{table}[htbp]
	\centering
	\caption{The scale corresponding to the different levels of the grid.}
	\label{gridscale}
	\begin{tabular}{lll}
		\hline
		level & The length of a grid(m) & The number of grids \\ \hline
		13    & 8000                    & 44                  \\
		14    & 4000                    & 147                 \\
		15    & 2000                    & 496                 \\
		16    & 1000                    & 1586                \\
		17    & 512                     & 6030                \\
		18    & 256                     & 23448               \\
		19    & 128                     & 92240               \\ \hline
	\end{tabular}
\end{table}

\subsubsection{Spatial autocorrelation}

(1) The Global Moran's Index statistic for spatial autocorrelation is given as \citep{KONG2021107344,KUMARI2019100239}:
\begin{equation}
	\label{GMorans}
	I = \dfrac{n}{S_{0}}\dfrac{\sum_{i=1}^{n}\sum_{j=1}^{n}w_{i,j}z_{i}z_{j}}{\sum_{i=1}^{n}z_{i}^{2}}
\end{equation}
where $z_{i}$ is the deviation of an attribute for feature $i$ from its mean, $w_{i,j}$ is the spatial weight between 
feature $i$ and $j$, $n$ equals the total numbers of features, and $S_{0}$ is the aggregate of all the spatial weights.

(2) The Local Morans's Index statistic of spatial association is given as \citep{TILLE2018182,TEPANOSYAN2019116}:
\begin{equation}
	\label{LMorans}
	I_{i} = \dfrac{x_{i}-\bar{X}}{S_{i}^{2}}\sum_{j=1,j\neq 1}^{n}w_{i,j}(x_{j}-\bar{X})
\end{equation}
where $x_{i}$ is an attribute for feature i, $\bar{X}$ is the mean of the corresponding attribute, 
$w_{i,j}$ is the spatial weight between feature $i$ and $j$, $n$ equals the total number of features, $z_{i}$ is the deviation of an attribute for feature $i$ from its mean.

The cluster/outlier type field distinguishes between a statistically significant cluster of high values (HH), cluster of low values (LL), cluster of not significant (NS), outlier in which a high value is surrounded primarily by low values (HL), and outlier in which a low value is surrounded primarily by high values (LH).

\section{Results}
\subsection{Changes in land use and ESV at the administrative districts level}
As shown in Table \ref{landuse}, in 2016, paddy filed occupied the most area in Haian City (39.13\%), followed by dryland (13.20\%), settlement (15.02\%), garden land (9.34\%), water area (9.20\%), structure (6.47\%), and grassland (4.71\%). However, the sum of the other two land use types, i.e., artificial pile Digging and forest, is less than 3\%. 

From 2016 to 2019, paddy field, dry land, garden land, grassland, and artificial pile Digging land showed negative growth, with the number of paddy field decreasing -1.78\%. 
Dryland, forest, settlement, structure, and water area showed an increasing trend, among which structure increased the most (+1.03\%), coinciding with the massive infrastructure investment in Haian in recent years.

\begin{table}[htbp]
	\centering
	\caption{Land use change in Haian City from 2016 to 2019}
	\label{landuse}
	\begin{tabular}{llll}
		\hline
		\multirow{2}{*}{Land over types}                                        & \multicolumn{2}{l}{Proportion} & Changed proportion \\ \cline{2-4} 
		& 2016    & 2019    &         \\ \hline
		Paddy field & 39.13\% & 37.35\% & -1.78\% \\ 
		Dry land    & 13.20\% & 13.80\% & 0.61\%  \\ 
		Garden land & 9.34\%  & 9.12\%  & -0.22\% \\ 
		Forest      & 1.87\%  & 2.41\%  & 0.55\%  \\ 
		Grassland   & 4.71\%  & 3.94\%  & -0.77\% \\ 
		Settlement  & 15.02\% & 15.52\% & 0.50\%  \\ 
		Structure   & 6.47\%  & 7.50\%  & 1.03\%  \\ 
		\begin{tabular}[c]{@{}l@{}}Artificial pile \\ digging land\end{tabular} & 1.04\%                  & 0.88\%                 & -0.16\%        \\ 
		Water area  & 9.20\%  & 9.47\%  & 0.27\%  \\ \hline
	\end{tabular}
\end{table}
We calculated the annual average ecosystem service value per unit area of different land types at urban scale according to the economic development status and the value coefficient of terrestrial ecosystem services in Haian City from 2016 to 2019. The total ESV values in 2016 and 2019 are 2,345 million RMB and 2,369 million RMB,  respectively, suggesting that the ecological value of Haian city in 2019 has increased by 24 million RMB compared with that in 2016. 

In 2016, the contribution rate of water area to the ESV was the highest (86.77\%), despite that the total area of water only accounted for 9.20\% of the whole study area. The contribution rates of paddy field, garden land, and grassland are 13.90\%, 10.84\%, and 6.96\%, respectively. The contribution rates for dryland, forest, and structure were low, only accounting for 3.97\%, 3.22\%, and 1.95\%, respectively. Negative contributions were found for settlement (-11.84\%) and artificial pile land (-0.08\%). From 2016 to 2019, the contribution rates of dry land, forest, structure, and water area have increased given the increased areas of these categories. The contribution of artificial pile Digging land has increased because of its deceased area and negative coefficient. Paddy filed, garden land, and water area contributed less in 2019 compared with 2016. In general, the ecological value of Haian City has increased from 2016 to 2019.

\begin{table}[htbp]
	\centering
	\caption{Overall change of ESV by land use type in Haian from 2016 to 2019}
	\resizebox{\textwidth}{!}{%
		\begin{tabular}{lllllll}
			\hline
			\multirow{2}{*}{Land over types} &
			\multicolumn{4}{l}{ESV(Million RMB )/Proportion (\%)} &
			\multirow{2}{*}{Value change} &
			\multirow{2}{*}{Proportional change(\%)} \\ \cline{2-5}
			& \multicolumn{2}{l}{2016} & \multicolumn{2}{l}{2019} &           &          \\ \hline
			Paddy field & 325.87    & 13.90\%    & 311.05    & 13.13\%    & -14.81 & -4.55\%  \\ 
			Dry land    & 93.16     & 3.97\%     & 97.44    & 4.11\%     & 4.28   & 4.60\%   \\ 
			Garden land & 245.77    & 10.48\%    & 23.99    & 10.13\%    & -5.81  & -2.37\%  \\ 
			Forest      & 75.43     & 3.22\%     & 97.47     & 4.11\%     & 22.04  & 29.22\%  \\ 
			Grassland   & 163.31    & 6.96\%     & 136.67    & 5.77\%     & -26.63 & -16.31\% \\ 
			Settlement  & -638.28   & -27.22\%   & -659.59   & -27.83\%   & -21.30 & 3.34\%   \\ 
			Structure   & 45.80     & 1.95\%     & 53.12     & 2.24\%     & 7.32   & 15.99\%  \\ 
			\begin{tabular}[c]{@{}l@{}}Artificial pile\\  Digging land\end{tabular} &
			            -0.79      &	-0.03\% &	-0.67 &	-0.03\% &		0.12 &	-15.19\% \\
			Water area  & 2,034.85   & 86.77\%    & 2,094.19   & 88.38\%    & 59.34  & 2.92\%   \\ 
			Total       & 2,345.13   & 100.00\%   & 2,369.67   & 100.00\%   & 24.53  & 17.65\%  \\ \hline
		\end{tabular}%
	}
\end{table}
In addition, we calculated the ESV of each administrative district of Haian City in 2016 and 2019, respectively, as shown in Figure \ref{fig:irgrid } below. Baidian District, Duntou District, Nanmo District, Dagong District have a relatively higher ESV than Qutang District, Yazhou District, Haian District, Chengdong District, and Libao District.
The ESV of city center was very low because it had a larger construction area.
\begin{figure}[htbp]
	\centering
	\includegraphics[scale=0.4]{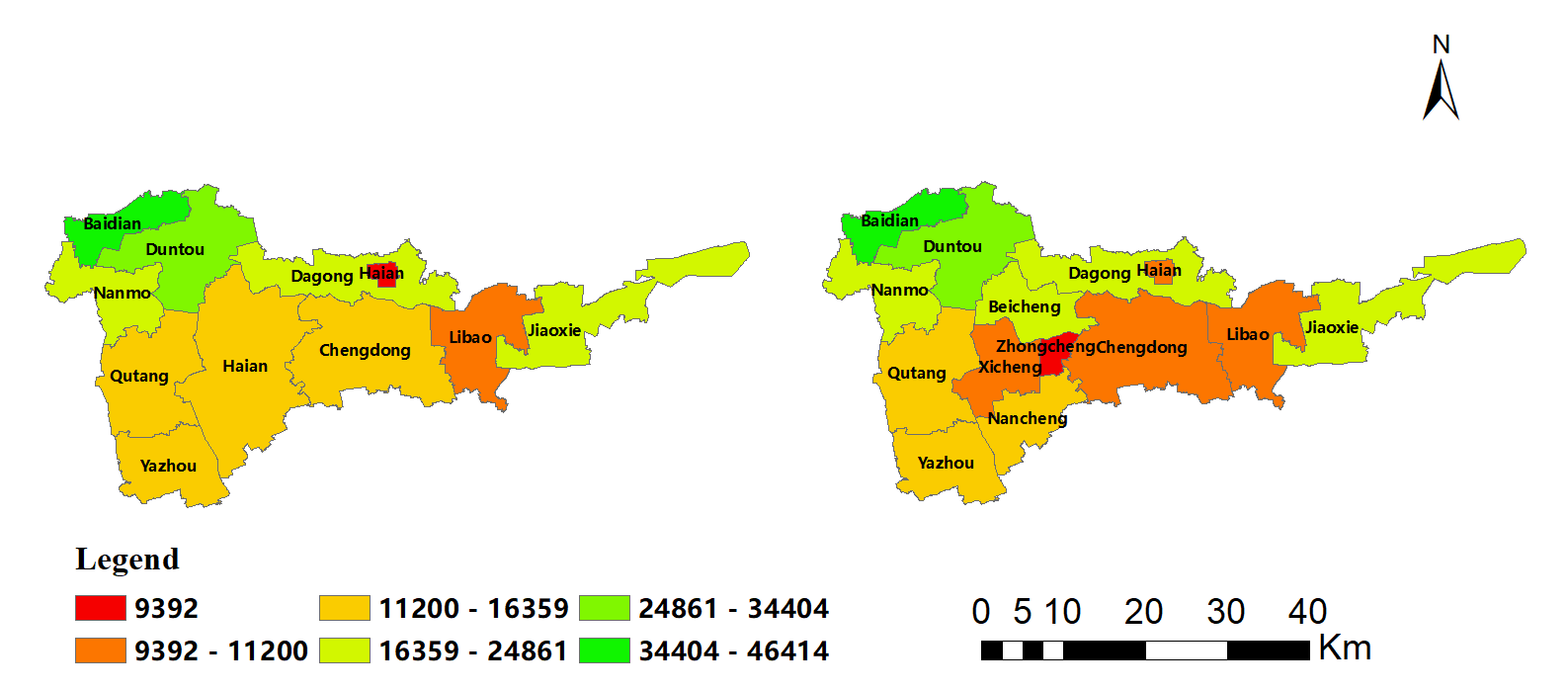}
	\caption{ESV change of Haian City by administrative district in 2016 and 2019}
	\label{fig:irgrid }
\end{figure}

\subsection{Monitoring ecological service value via dynamic multi-level grids}

\begin{figure}[htbp]
	\centering
	\subfigure[grid14]{\includegraphics[scale=0.19]{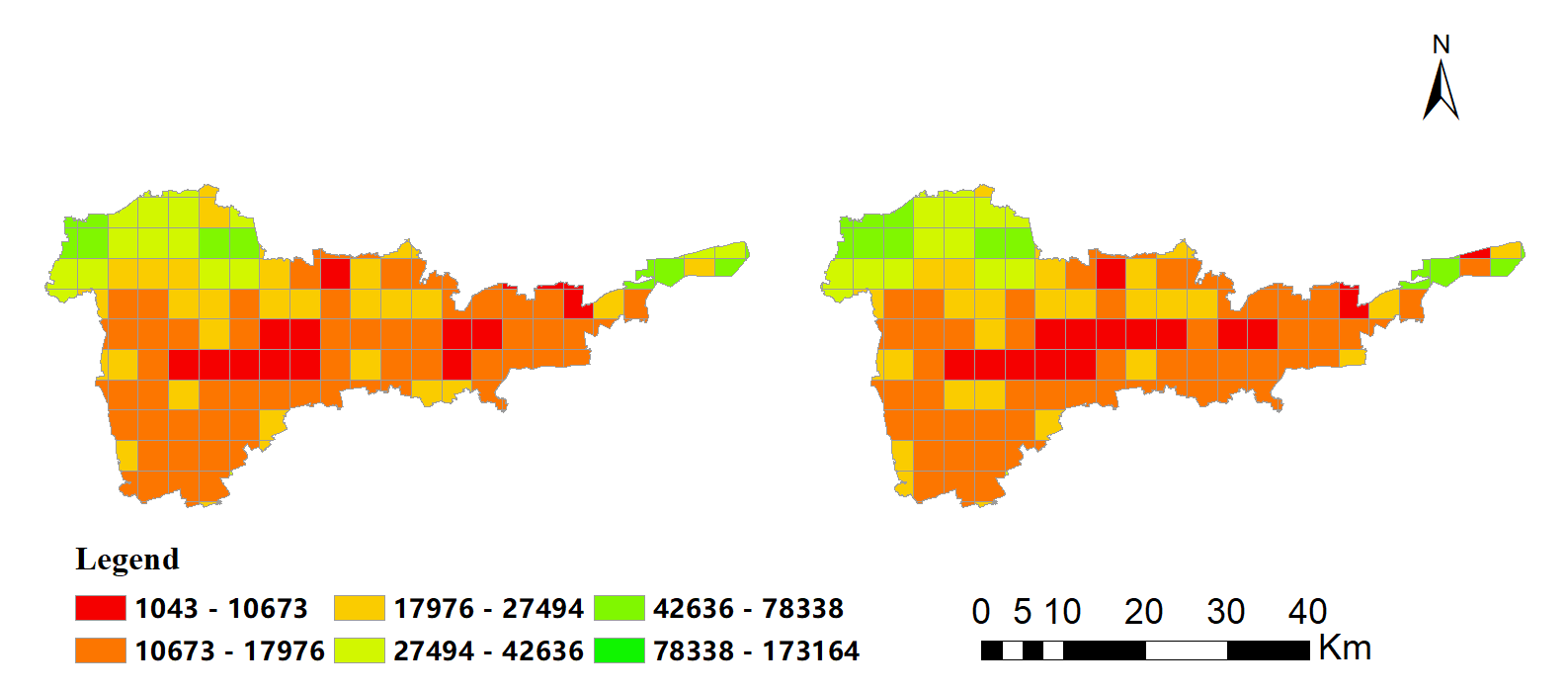}}
	\subfigure[grid15]{\includegraphics[scale=0.19]{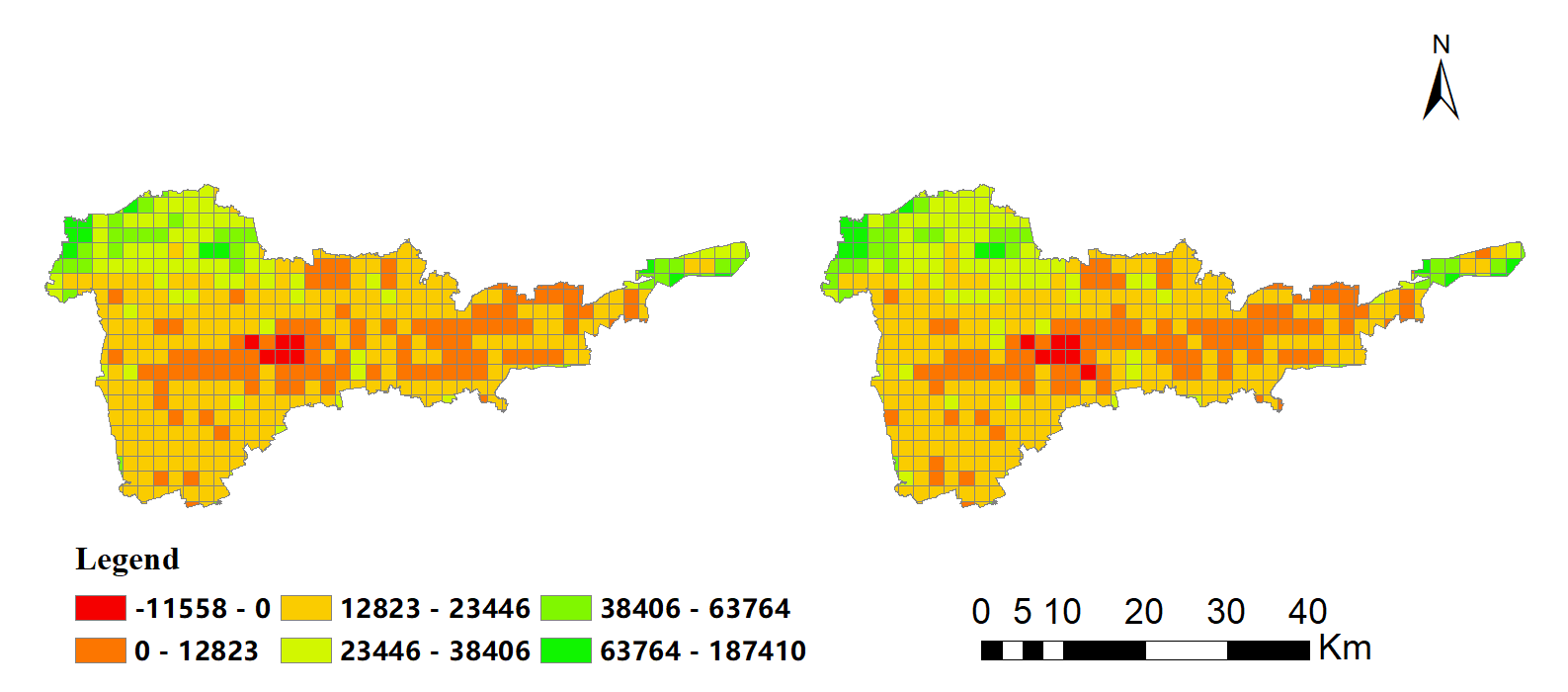}}
	\subfigure[grid16]{\includegraphics[scale=0.19]{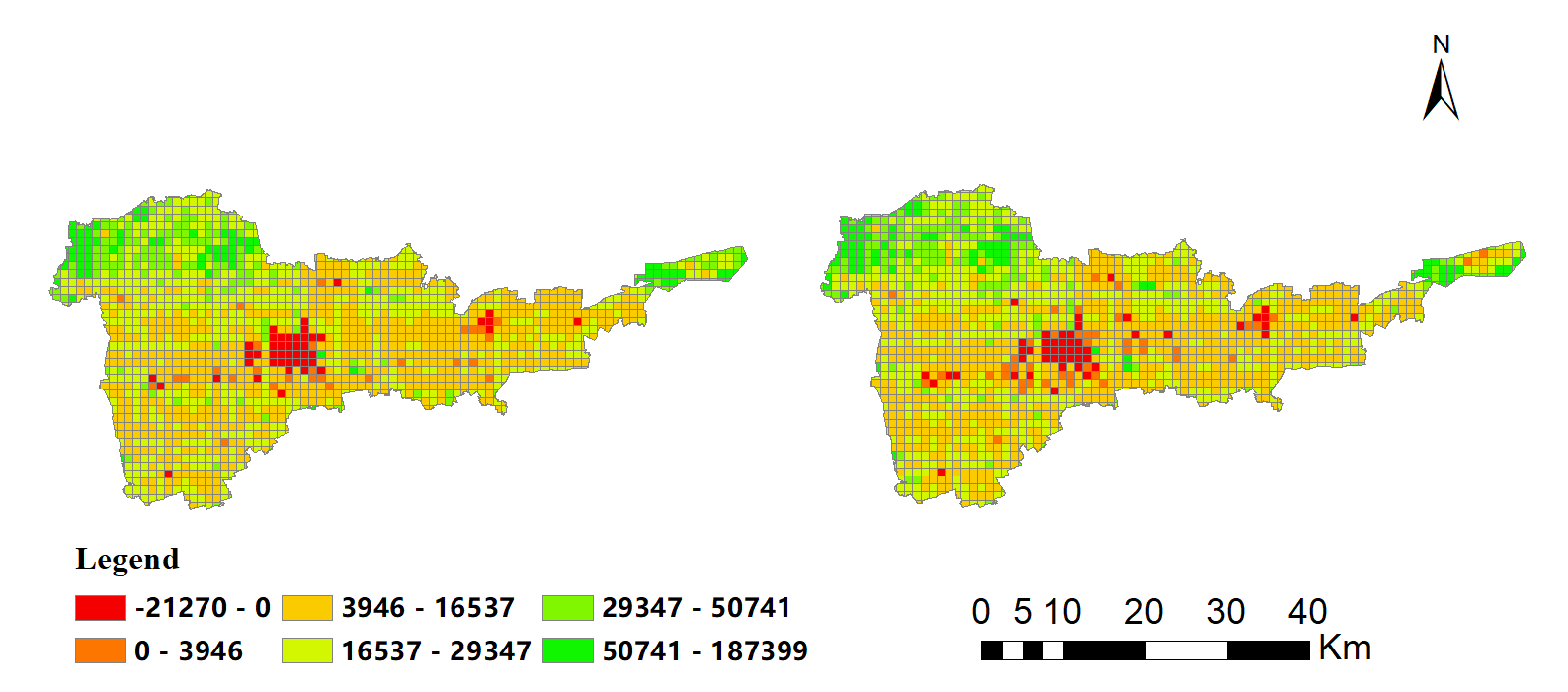}}
	\subfigure[grid17]{\includegraphics[scale=0.19]{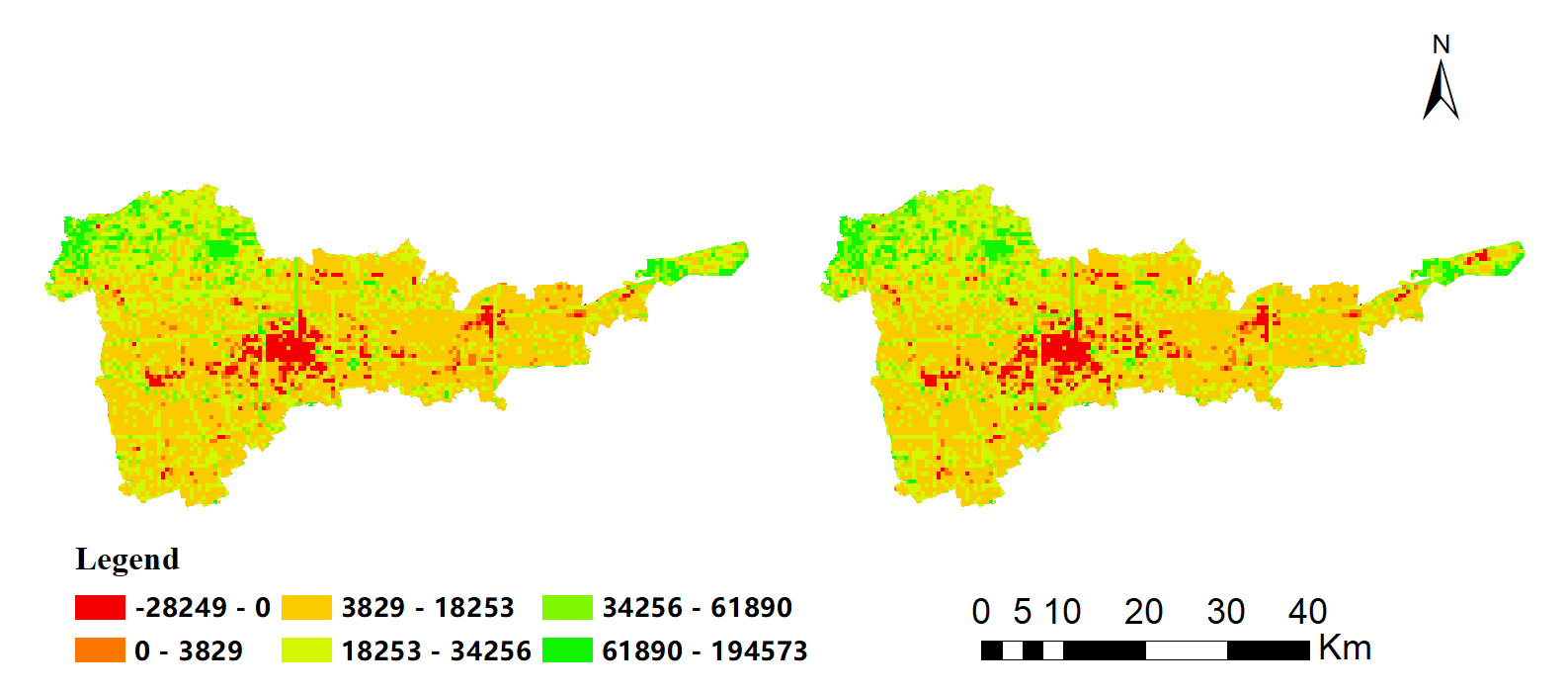}}
	\subfigure[grid18]{\includegraphics[scale=0.19]{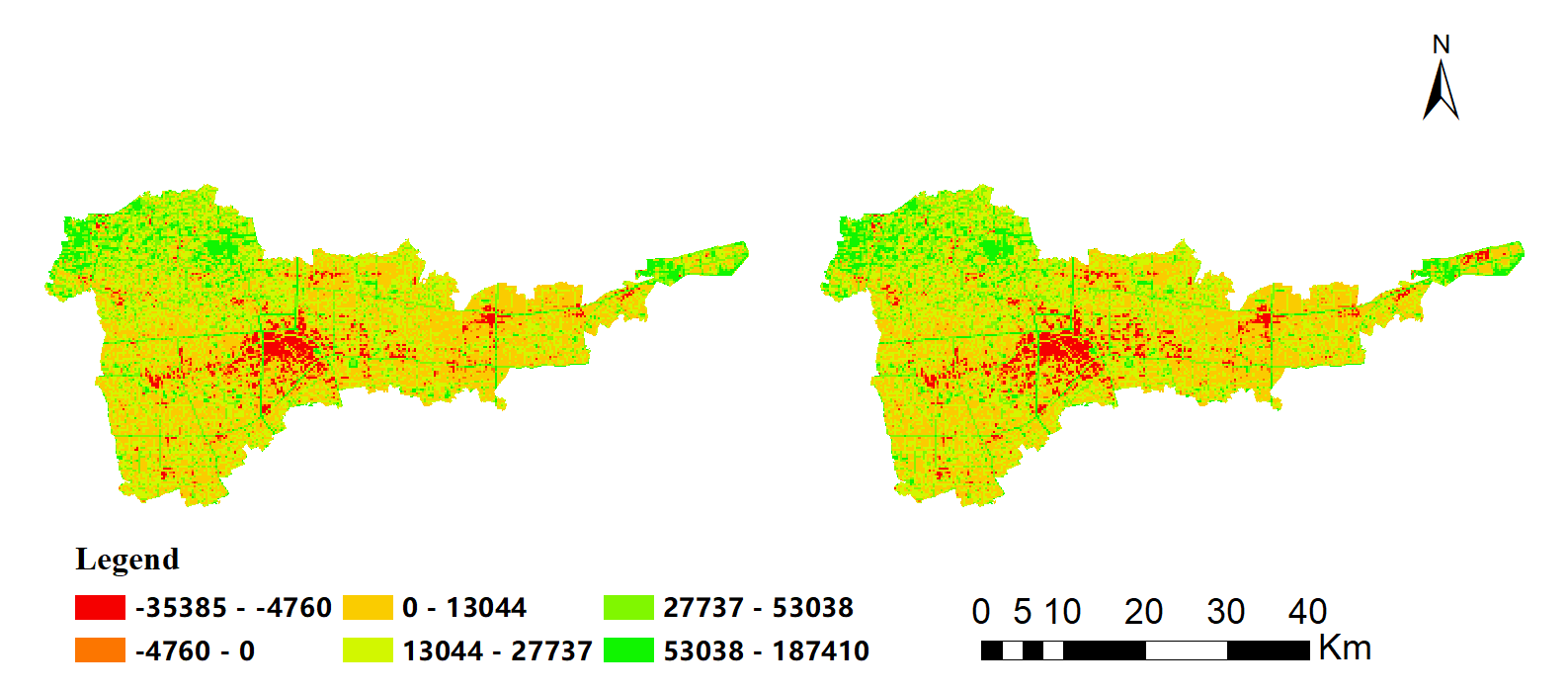}}
	\subfigure[grid19]{\includegraphics[scale=0.19]{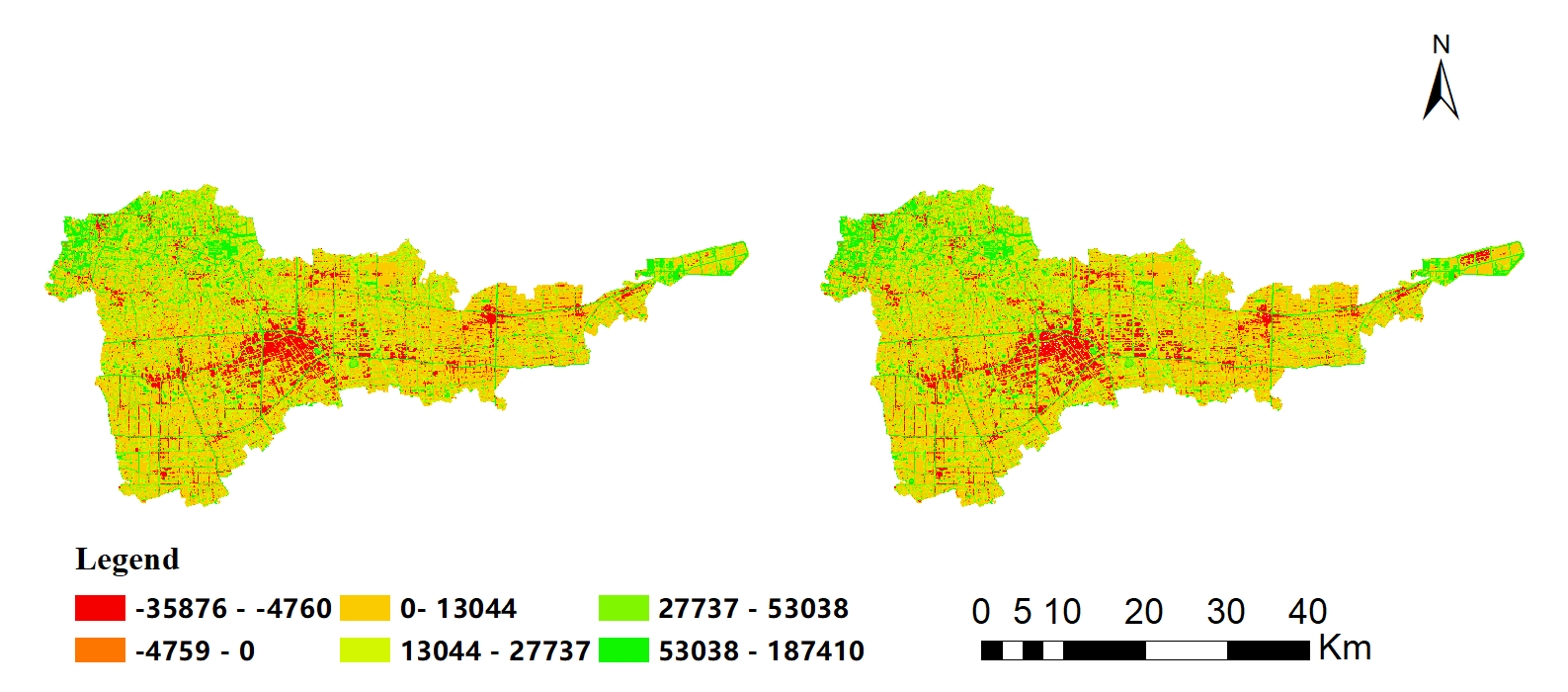}}
	\caption{ESV distribution of the grid levels 14-19 in 2016 and 2019 in Haian City.}
	\label{fig:grid}
\end{figure}

\begin{figure}[htbp]
	\centering
	\subfigure[grid14]{\includegraphics[scale=0.25]{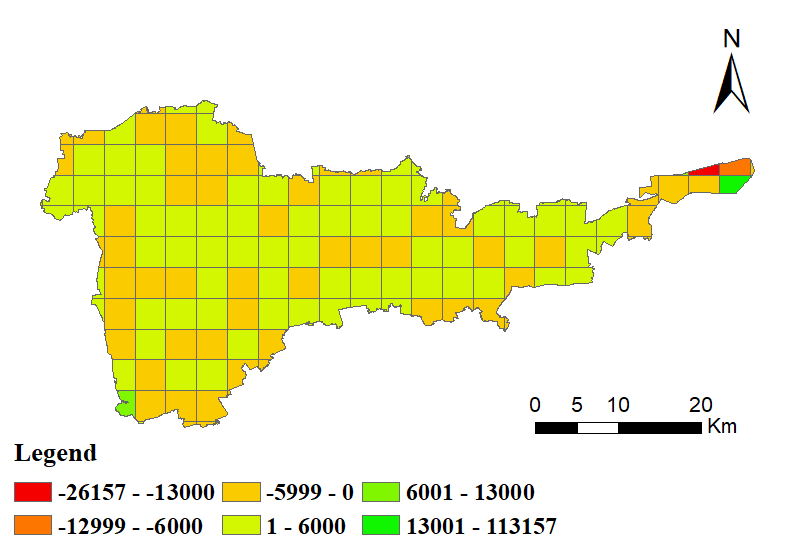}}
	\subfigure[grid15]{\includegraphics[scale=0.25]{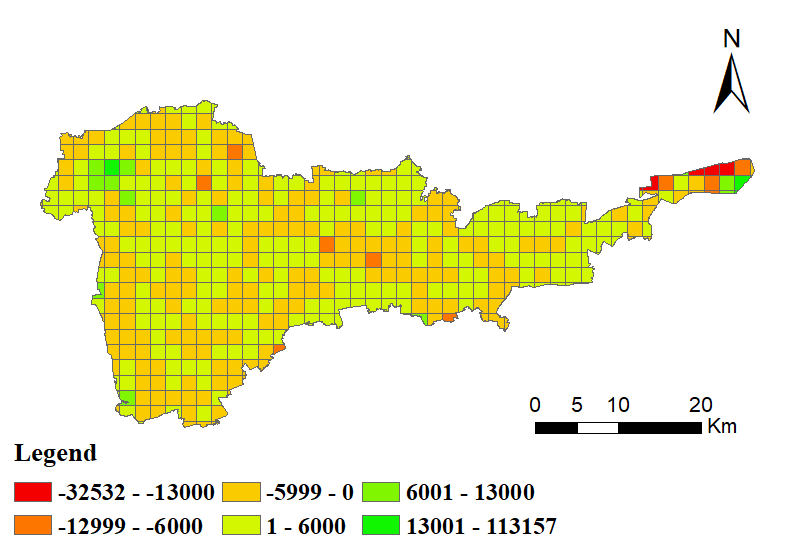}}
	\subfigure[grid16]{\includegraphics[scale=0.25]{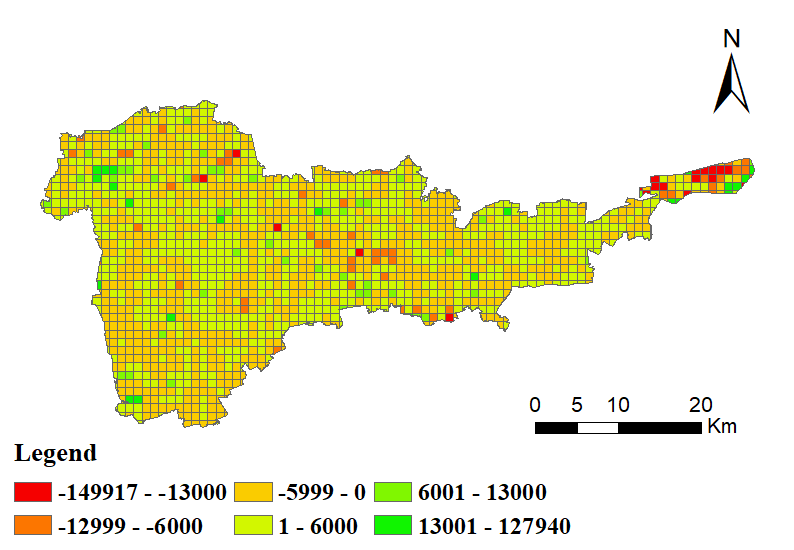}}
	\subfigure[grid17]{\includegraphics[scale=0.25]{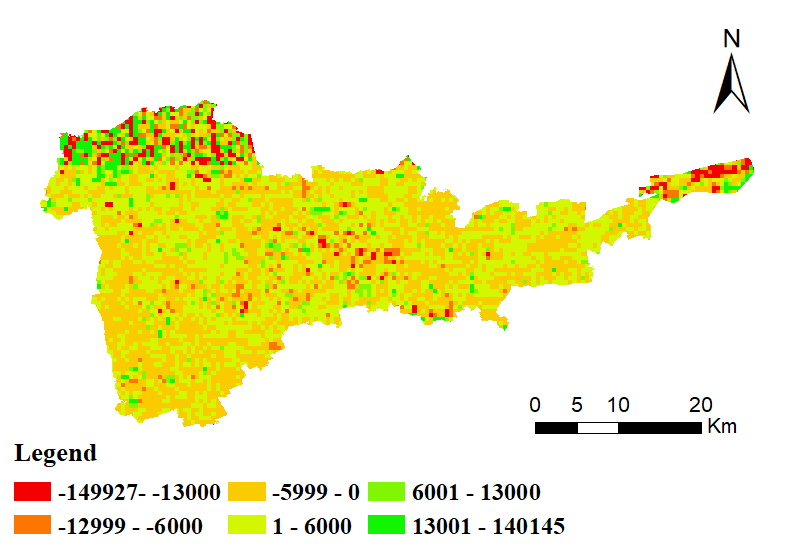}}
	\subfigure[grid18]{\includegraphics[scale=0.25]{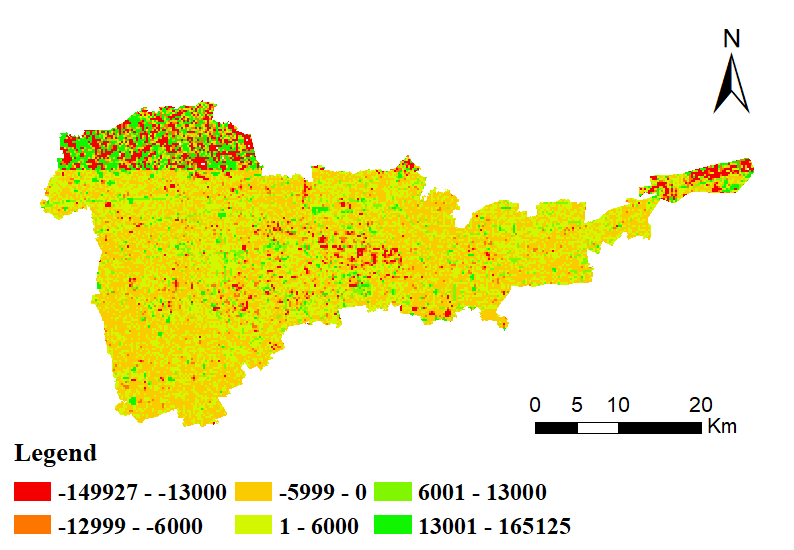}}
	\subfigure[grid19]{\includegraphics[scale=0.25]{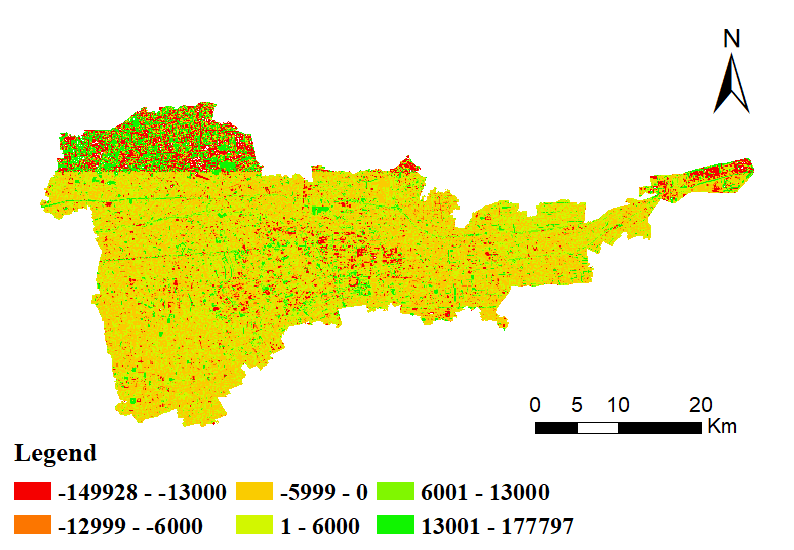}}
	\caption{Change of ESV at grid levels 14-19 from 2016 to 2019 in Haian City.}
	\label{fig:gridchange}
\end{figure}

ESV calculation based on administrative districts emphasizes more on revealing macro-scale ecological values. However, the multi-level grid-based ecological value is able to reveal the spatial dependence among geographical features and focuses on reflecting the impact on urban ecological value from different various scales. Spatial dependency is typically constrained by the observation scale, usually to evaluate the spatial determinants of urban ESV distribution based on different spatial scales. For ESV of grid cells, we calculated the ESV of 14-19 grids in Haian City in 2016 and 2019, given our proposed formula \ref{ESVg}.
Compared with the division based on administrative regions, the distribution of ESV at 14-16 grid levels reveal the multi-level characteristics of ESV in Haian City (Table \ref{fig:grid}). We observed that low values of ESV were concentrated in the center of the city, and the ESV values gradually increased outward. Detailed spatial distribution can be seen from the grid levels 17-19, where rivers can be clearly seen in green lines. The center of the city had more grids with low ESV values (red grids) in 2019 compared with 2016, which was in line with the urban expansion in the center of the city during the three year period.
We also calculated the change value of ESV from 2016 to 2019 as shown in the Figure \ref{fig:gridchange}.
In the study area, the value of ecological services increased from 2016 to 2019, indicating that the ecological environment level of the study area did not decline during the recent rapid economic development.

\subsection{Coefficient of sensitivity (CS) analysis for ecosystem service value (ESV) in grid}
Since the ecological value coefficient per unit area of various land types is allocated based on the most similar equivalent ecosystem, meaning that these equivalent ecosystems do not completely match with the land types they represent, thus leading to certain uncertainty in the valuation of ESV. Therefore, we conduct additional sensitivity analysis to test the dependence of ESV obtained in this study on the change of value coefficient.

\begin{figure}[htbp]
	\centering
	\includegraphics[scale=0.21]{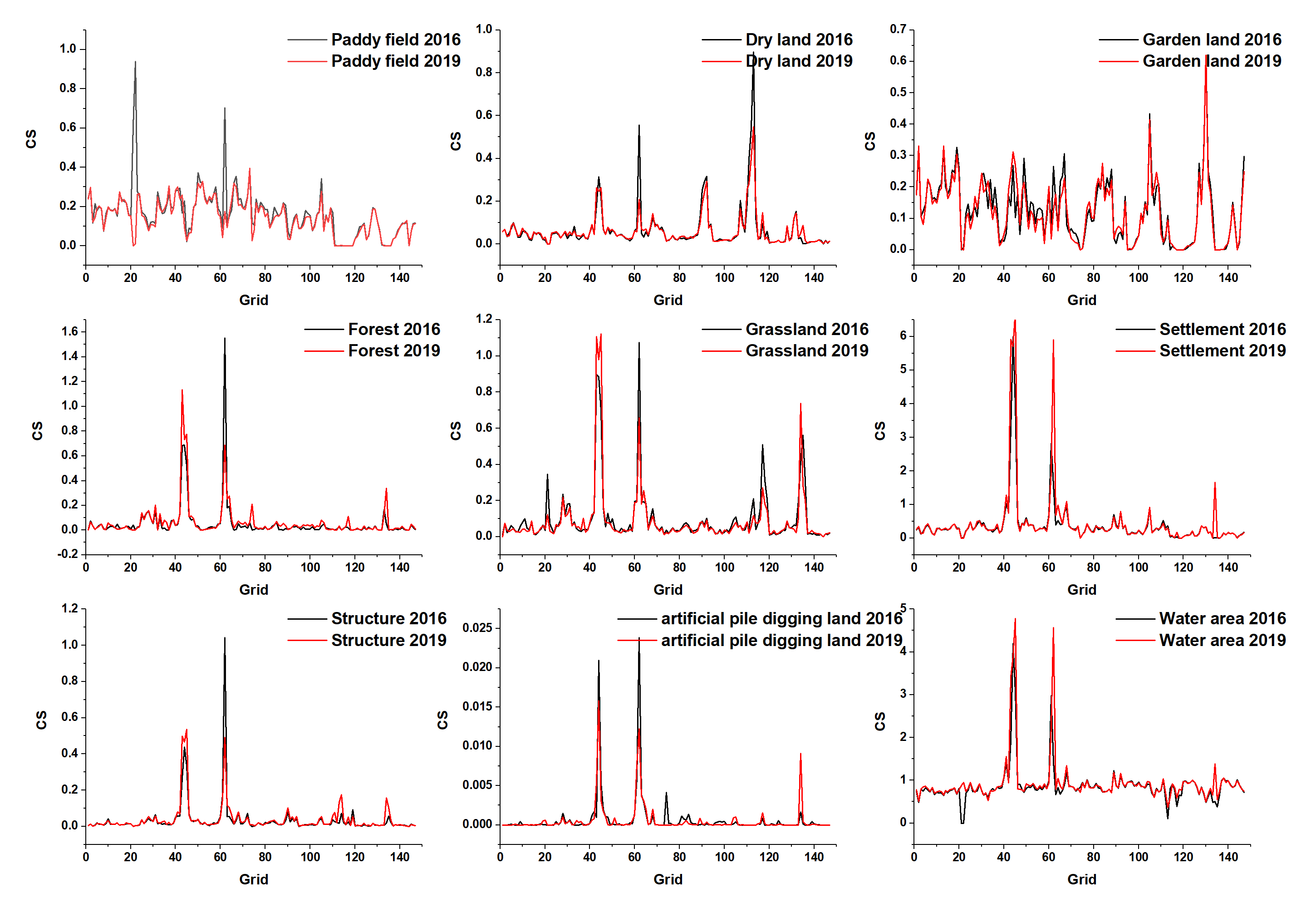}
	\caption{Percentage change in estimated total ESV and coefficients of sensitivity (CSs) from adjustment of ecosystem valuation coefficients (VC).}
	\label{fig:CS}
\end{figure}

According to formula \ref{CS}, we calculated the CS of the whole ESV in the study area, as shown in Table \ref{table:cs} and Figure \ref{fig:CS}. 
It illustrates that the CS of all land use types is less than 1, implying that the ESV measured in this study was inelastic to the value coefficient, meaning that the ESV values obtained in this study are accurate, despite some uncertainty in the value coefficients. Among all land use types, the sensitivity of water was found much higher than that of other land types, mainly due to its extensive spatial coverage and highest value coefficient.

\begin{table}[htbp]
	\centering
	\caption{Percentage change in estimated total ESV and coefficients of sensitivity (CSs) from adjustment of ecosystem valuation coefficients (VC).}
	\label{table:cs}
		\begin{tabular}{lllll}
			\hline
			\multirow{2}{*}{Change of value coefficient}                                      & \multicolumn{2}{l}{2016} & \multicolumn{2}{l}{2019} \\ \cline{2-5} 
			& ESV\%   & CS     & ESV\%   & CS     \\ \hline
			Paddy field VC $\pm$ 50\% & 6.95\%  & 0.1390 & 6.56\%  & 0.1313 \\
			Dry land VC $\pm$ 50\%    & 1.99\%  & 0.0397 & 2.06\%  & 0.0411 \\
			Garden land VC $\pm$ 50\% & 5.24\%  & 0.1048 & 5.06\%  & 0.1013 \\
			Forest VC $\pm$ 50\%      & 1.61\%  & 0.0322 & 2.06\%  & 0.0411 \\
			Grassland VC $\pm$ 50\%   & 3.48\%  & 0.0696 & 2.88\%  & 0.0577 \\
			Settlement VC $\pm$ 50\%  & 13.61\% & 0.2722 & 13.92\% & 0.2783 \\
			Structure VC $\pm$ 50\%   & 0.98\%  & 0.0195 & 1.12\%  & 0.0224 \\
			\begin{tabular}[c]{@{}l@{}}Artificial pile \\ Digging land VC $\pm$ 50\%\end{tabular} & 0.02\%      & 0.0003     & 0.01\%      & 0.0003     \\
			Water area VC $\pm$ 50\%  & 43.38\% & 0.8677 & 44.19\% & 0.8838 \\ \hline
		\end{tabular}%
\end{table}

\section{Discussion}

\subsection{Dynamic multi-level grids-based spatial autocorrelation analysis of urban ESV}

\begin{figure}[htbp]
	\centering
	\subfigure[2016]{\includegraphics[scale=0.3]{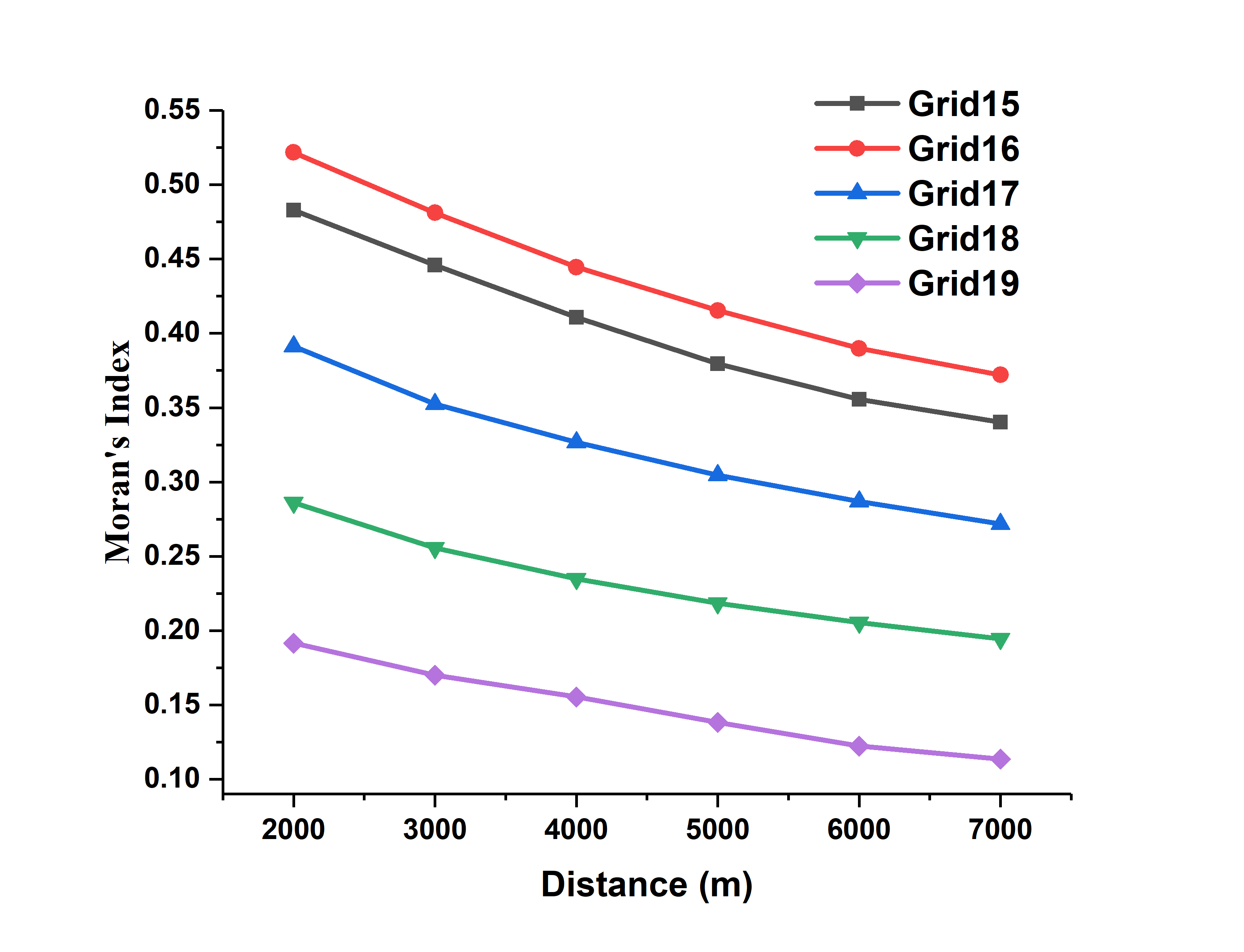}}
	\subfigure[2019]{\includegraphics[scale=0.3]{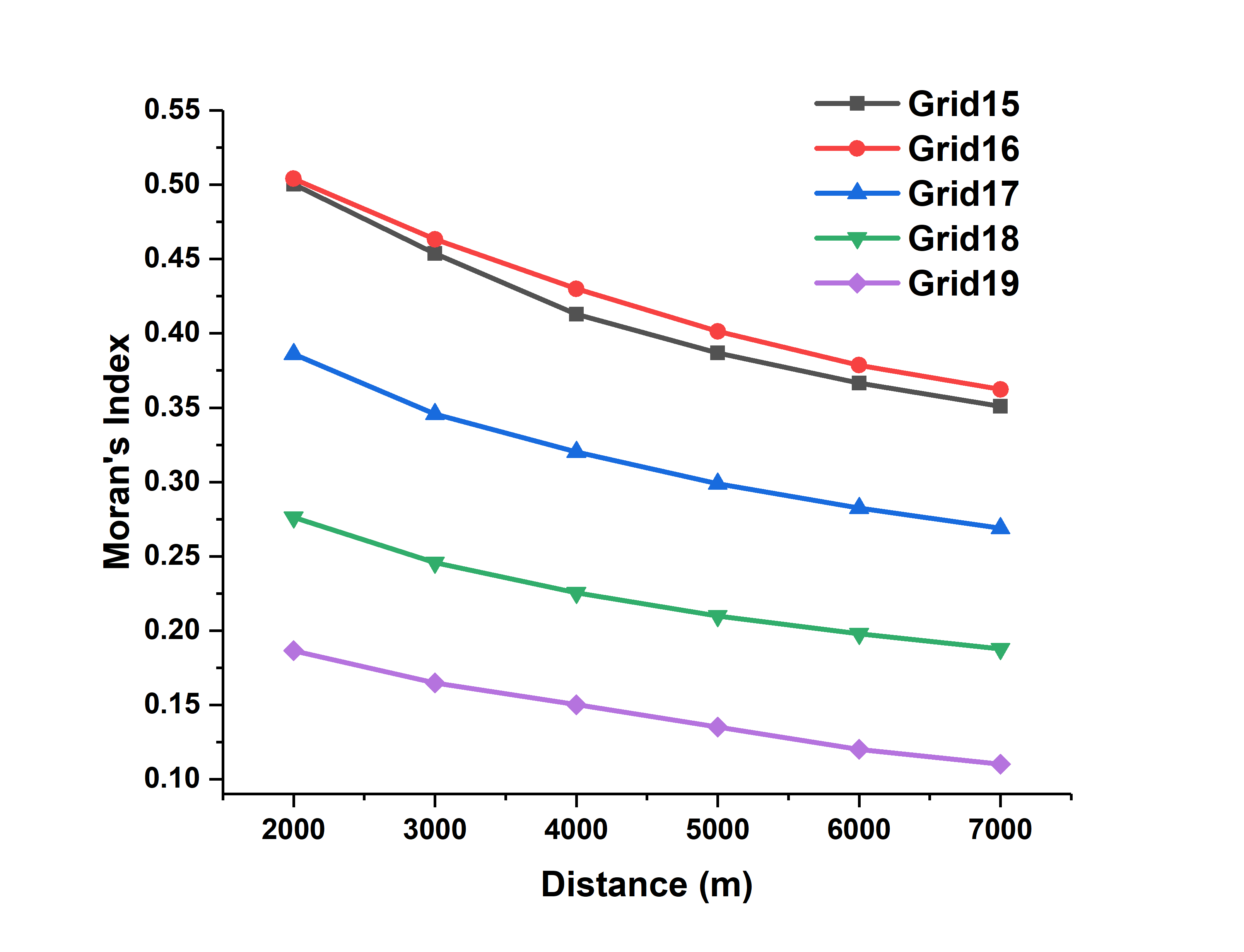}}
	\caption{Change of ESV from grid levels 14-19 in 2016 and 2019 in Haian City.}
	\label{fig:morans}
\end{figure}

This study analyzed the spatial autocorrelation characteristics of the urban ESV in Haian City on five scales using the Global Moran’s Index (Formula \ref{GMorans}), to analyze information hidden in the scale.
When calculating the Global Moran’s Index of urban ESV in five spatial scales, we divided the distances of 2000 m, 3000 m, 4000 m, 5000, 6000 m, and 7000 m to ensure that each grid had at least one neighbor.
Figure \ref{fig:morans} showed that urban ESV exhibited a strong spatial correlation on all five scales, and as the distance increases, the spatial autocorrelation gradually weakens, which conformed to Tobler's First Law of Geography.
What’s more, our results suggested that
the Globe Moran’s Index value gradually decreases with the increase of the spatial scale, except that the value of grid16 is greater than the value of grid15.
 Since the threshold distance used in our experiment was longer for fine grids than for rough grids, the Globe Moran's Index value gradually decreases under the same threshold distance.
In addition, there was not much difference in the fine grid compared to the two-year scale, but the difference in the rough grid was obvious.
The autocorrelation of urban ESV was higher than that of Moran’s Index on a larger scale, indicating that urban expansion was more regionalized.

\begin{figure}[htbp]
	\centering
	\subfigure[grid14]{\includegraphics[scale=0.19]{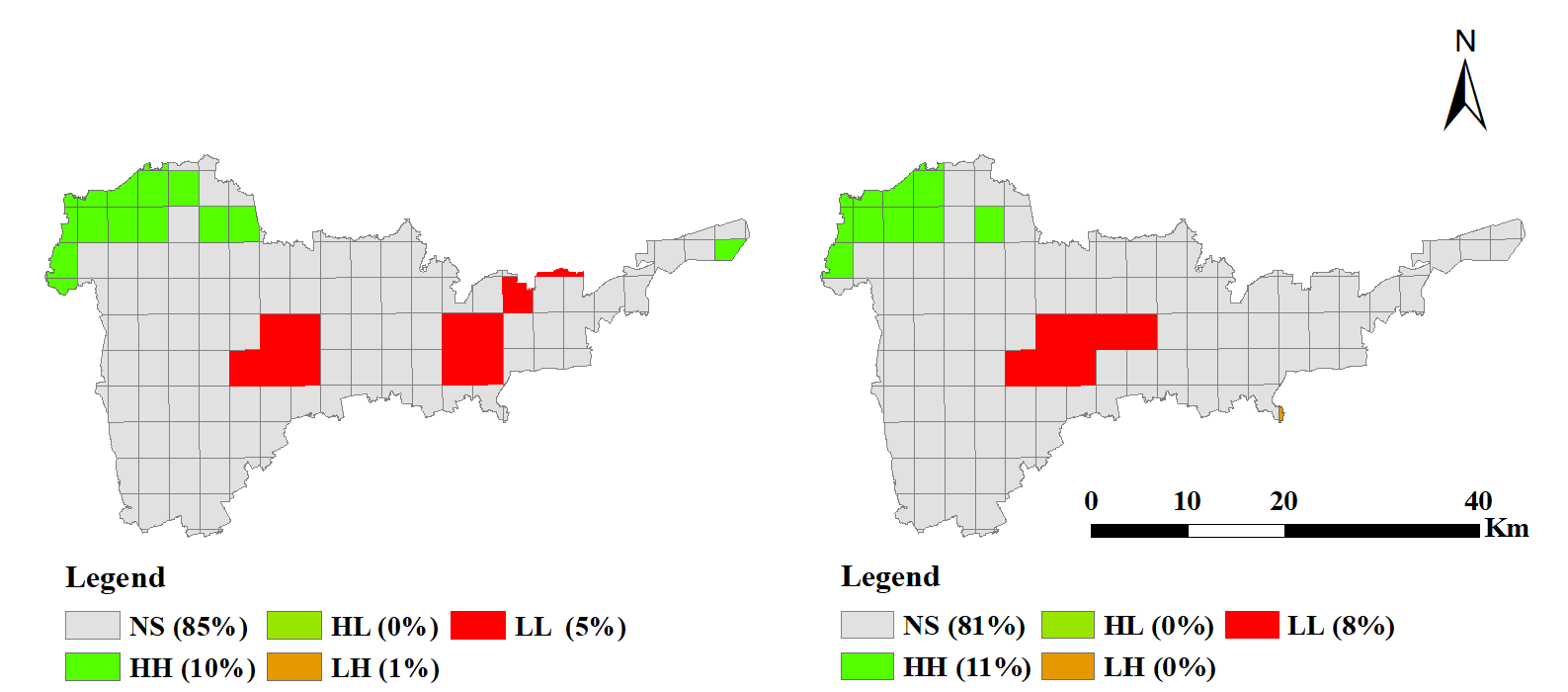}}
	\subfigure[grid15]{\includegraphics[scale=0.19]{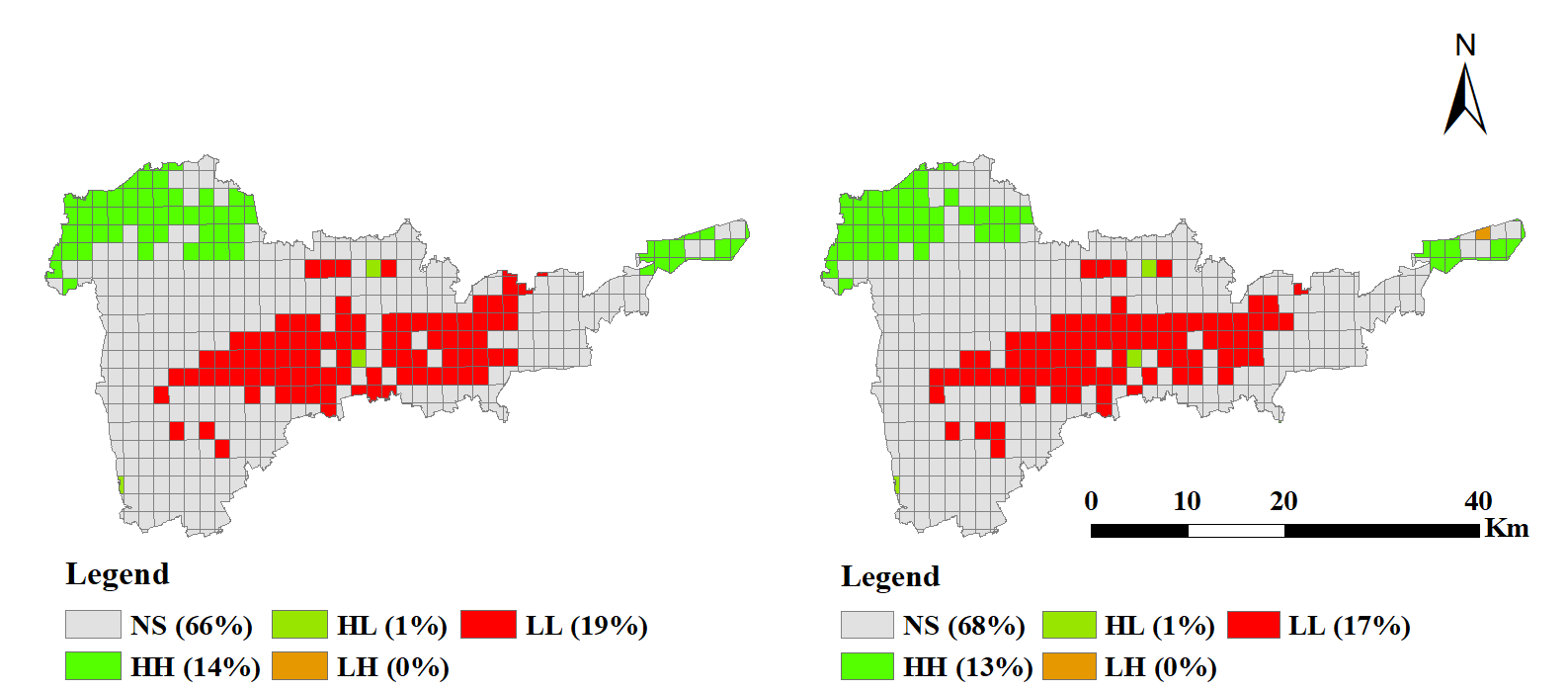}}
	\subfigure[grid16]{\includegraphics[scale=0.19]{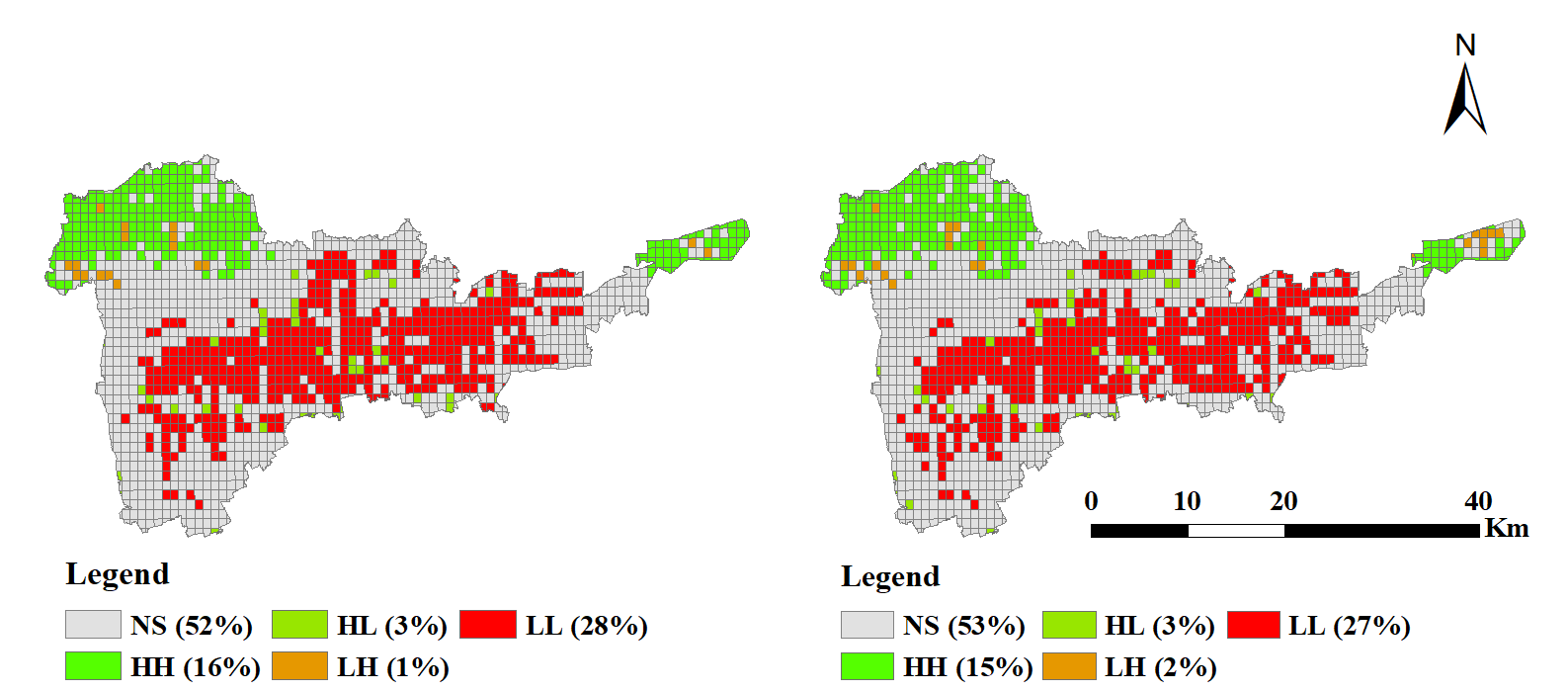}}
	\subfigure[grid17]{\includegraphics[scale=0.19]{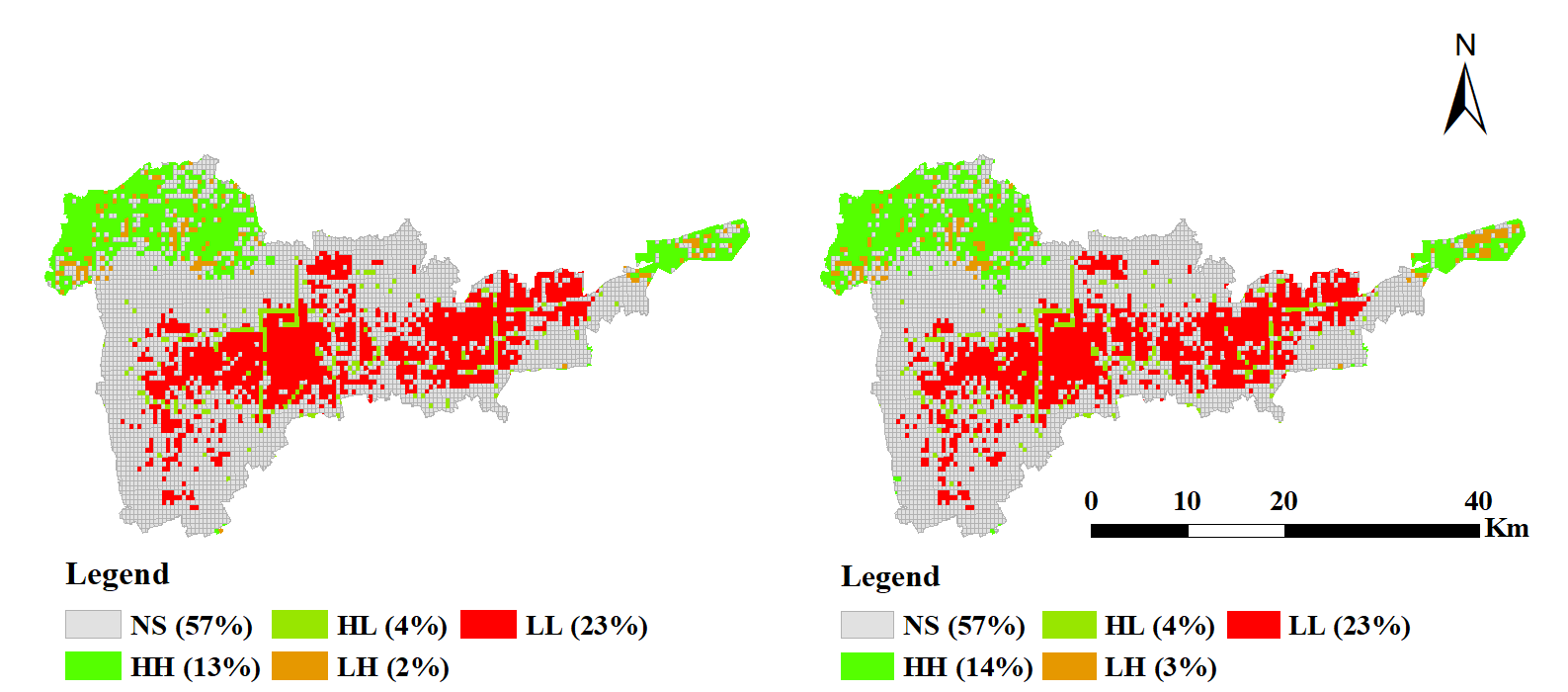}}
	\subfigure[grid18]{\includegraphics[scale=0.19]{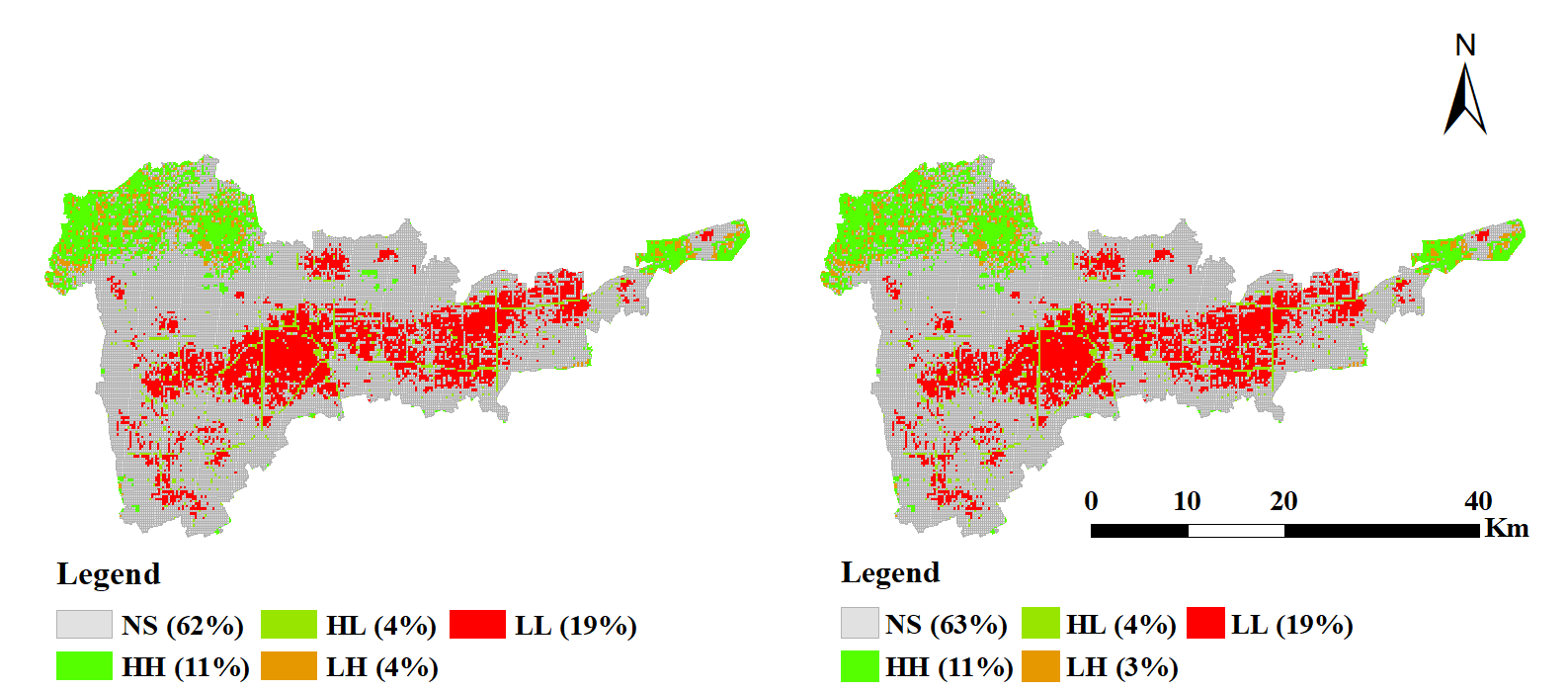}}
	\subfigure[grid19]{\includegraphics[scale=0.19]{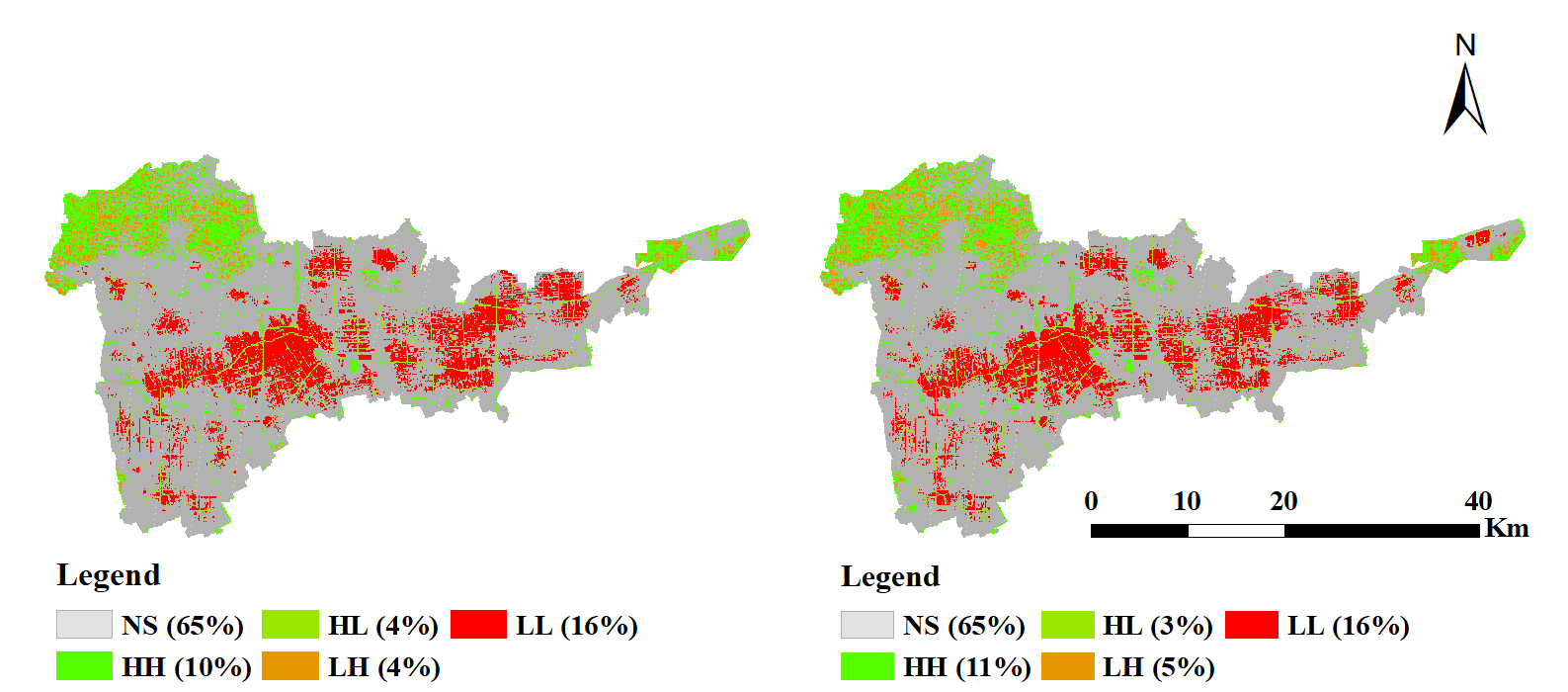}}
	\caption{ESV Cluster and Outlier Analysis of the grid levels 14-19 in 2016 and 2019 in Haian City.}
	\label{fig:morans_cluster}
\end{figure}

The results of local Moran's Index(Formula \ref{LMorans}) are displayed in 2016 and 2019 (Figure \ref{fig:morans_cluster} ). 
The ESV in Haian was mainly composed of HH and LL. The LL aggregation areas were mostly found in the northwestern, southern-central, and  eastern regions. In comparison, the HH aggregation areas were mainly located in the northern regions as well as partly in the western areas. 
Compared to 2016, there was no specific improvement of the clusters' number  in the central part of Haian City in 2019, and there was a large concentration of LH in the north-eastern part of the city.
Our results well demonstrated the superiority of DMLG. 
From the 14-level to the 19-level, the area of the LL aggregation in the northwestern part decreased and was mainly turned into HH aggregation, indicating the improvement of the ecological environment in these regions. 
In the central and eastern parts of Haian City, however, many isolated LL regions were gradually aggregated.
This was because there are more grids on the fine scale, and it was easier to display the clustering properties of the ESV than the rough grids.
The proportion of LL, HH, LH, HL in grids from 14 to 19 levels tends to increase first and then decrease, suggesting that as the fineness of the grid rises, the number of clusters increases and the precision of clustering improves.

\subsection{Comparison between the multilevel grid-based methods and existing methods}

\subsubsection{Strengths of the dynamic multi-level grids-based (DMLG) methods}

We compared our approach against four typical methods for evaluating ecological service value in terms of research content, evaluation unit, evaluation method (Table \ref{comMethods}) \cite{LI2021106873}. 
The advantages of using dynamic multi-level grids in evaluating ESV are obvious. 

Most scholars such as \citet{sunSpatiotemporalEvolutionScenarios2019,zhangSpatialRelationshipEcosystem2018,yongxiuSpatiotemporalVariationsCoupling2020} used macro spatial scales that include: urban agglomerations, cities, townships, etc.
This macro-perspective ignores the spatial heterogeneity within cities, leading to a variety of problems such as the failure to identify certain special, important, and subtle ecological phenomena.
Some researchers tried to increase the spatial scale of ESV based on remote sensing images by selecting the smallest unit of remote sensing image, namely pixel, aiming to reflect regional heterogeneity \citep{LI2021106873}.
However, selecting the pixels as the spatial scale fails to explore the spatial heterogeneity at multi-level scales. Filling these gaps, we propose a dynamic multi-level grids-based (DMLG) ESV calculation method. 

Taking the finest 19-level grid as an example, each grid unit is 128m by 128m, finer than the smallest administrative unit (community level). In addition, the visualization and details of dynamic multi-level grid-based results increase as the grid level increases, providing a flexible microscopic spatial scale and uncovering the information hidden under the administrative scale. Although no feasible method has been proposed to verify the assessment accuracy of ESV, the validity of ESV can still be indirectly reflected through accurate verification of relevant parameters, such as land cover classification and sensitivity coefficient. The conceptual, methodological, and experimental knowledge in this study is expected to largely benefit the fine-grained monitoring of urban ecology.

\begin{table}[htbp]
	\centering
	\caption{ Comparison between the DMLG and existing methods.}
    \label{comMethods}
	\resizebox{\textwidth}{!}{%
		\begin{tabular}{lllll}
			\hline
			\begin{tabular}[c]{@{}l@{}}Literature\\ source\end{tabular} &
			\begin{tabular}[c]{@{}l@{}}Study \\ area\end{tabular} &
			Research content &
			\begin{tabular}[c]{@{}l@{}}Evaluation \\ unit\end{tabular} &
			\begin{tabular}[c]{@{}l@{}}Evaluation\\ method\end{tabular} \\ \hline
			\citep{sunSpatiotemporalEvolutionScenarios2019} &
			China &
			\begin{tabular}[c]{@{}l@{}}Spatio-temporal \\ evolution scenarios \\ and the coupling \\ analysis of ecosystem \\ services\end{tabular} &
			Administrative region &
			\begin{tabular}[c]{@{}l@{}}LULC matrix \\ approach\end{tabular} \\
			\citep{zhangSpatialRelationshipEcosystem2018} &
			\begin{tabular}[c]{@{}l@{}}Wuhan, \\ Hubei \\ province,\\ China\end{tabular} &
			\begin{tabular}[c]{@{}l@{}}On the spatial \\ relationship between \\ ecosystem services \\ and urbanization\end{tabular} &
			Landscape   types &
			\begin{tabular}[c]{@{}l@{}}Spatial \\ regression \\ models\end{tabular} \\
			\citep{LI2021106873} &
			\begin{tabular}[c]{@{}l@{}}Yanzhou \\ coalfield,\\ Shandong \\ Province, \\ China\end{tabular} &
			\begin{tabular}[c]{@{}l@{}}Effects of \\ urbanization \\ on ecosystem \\ service values \\ in a mineral \\ resource-based city\end{tabular} &
			Pixels &
			\begin{tabular}[c]{@{}l@{}}Pixel-based\\ time series\\ model of ESV\end{tabular} \\
			\citep{yongxiuSpatiotemporalVariationsCoupling2020} &
			\begin{tabular}[c]{@{}l@{}}Qinghai-Tibet \\ Plateau, \\ China\end{tabular} &
			\begin{tabular}[c]{@{}l@{}}Spatio-temporal \\ variations and \\ coupling \\ of human activity \\ intensity and \\ ecosystem services\end{tabular} &
			Administrative region &
			\begin{tabular}[c]{@{}l@{}}four-quadrant\\ model\end{tabular} \\
			Ours &
			\begin{tabular}[c]{@{}l@{}}Haian,\\ Jiangsu, \\ China\end{tabular} &
			\begin{tabular}[c]{@{}l@{}}Urban ecosystem \\ service value \\ monitoring\end{tabular} &
			grid &
			\begin{tabular}[c]{@{}l@{}}DMLG \\ regular \\ grid\end{tabular} \\ \hline
		\end{tabular}%
	}
\end{table}
\subsubsection{Limitations}
Although the DMLG reveals the spatio-temporal dynamic characteristics of ESV, we need to acknowledge its limitations. 
The accuracy of value coefficient of ESV directly affects the accuracy of our results. Although no feasible method has been proposed to verify the assessment accuracy of ESV, the validity of ESV can still be indirectly reflected through accurate verification of relevant parameters, such as land cover classification, and sensitivity coefficient. 
Based on the collection of several papers and related time and space coefficients, we try to make it precise.

In addition, the DMLG method takes vector data as input for calculation, leading to a high amount of computation when massive grids need to be calculated.
Due to the huge computing demand at high-level grids, we set the most detailed scale as 19-level grids, with the length of each grid cell 128m. The level of details and visualization of ESV improves as the level increases.

\section{Conclusion}
In this paper, taking Haian City as the research area, we propose DMLG ESV evaluation method, while calculating ESV using traditional administrative scales. The results confirm the effectiveness of our method. Conclusion as follows:

(1) From 2016 to 2019, the ESV of the study area increased by 24.54 million RMB. The negative values in ESV are concentrated in the central urban area, which shows a trend of outward expansion, indicating that the ongoing urban expanse does not reduce the ecological growth in the study area. 

(2) Our method not only improves the limitations of the existing macro-scale models, but also has multi-level functions.  Grids in lower levels (rough) can provide macroscopic decision analysis, while Grids in higher levels (fine) can be used to capture subtle spatial heterogeneity. 

(3)	The application of DMLG in this study to evaluate the city-level ecological value can overcome the limitations of the existing methods, which often ignore the varying spatial heterogeneity by implementing a fixed spatial unit.

Proposals on sustainable development should take full account of local development needs and select appropriate ecological optimization measures.

%
%

\bibliographystyle{cas-model2-names}

\bibliography{grid-paper}





\end{document}